\documentclass[superscriptaddress,showpacs,aps,twocolumn,prb,floatfix]{revtex4}

\usepackage{amssymb}
\usepackage{amsmath}
\usepackage[dvips]{graphicx}
\usepackage{subfigure}
\usepackage{bm}
\usepackage{color}
\usepackage{gensymb}
\usepackage{wasysym}
\usepackage{xcolor,soul}
\usepackage{url}

\begin{document}
 
\title{Magnetic transition and spin-polarized two-dimensional electron gas controlled by polarization switching in strained CaMnO$_3$/BaTiO$_3$ slabs}

\author{S. Di Napoli}
\affiliation{Instituto de Nanociencia y Nanotecnolog\'{\i}a (CNEA - CONICET), Nodo Buenos Aires, Av. Gral. Paz 1499, B1650 Villa Maip\'u, Provincia de Buenos Aires, Argentina}
\affiliation{Departamento de F\'{\i}sica de la Materia Condensada, GIyA-CNEA, Av. Gral. Paz 1499, B1650 Villa Maip\'u, Provincia de Buenos Aires, Argentina}

\author{A. Rom\'an}
\affiliation{Instituto de Nanociencia y Nanotecnolog\'{\i}a (CNEA - CONICET), Nodo Buenos Aires, Av. Gral. Paz 1499, B1650 Villa Maip\'u, Provincia de Buenos Aires, Argentina}
\affiliation{Laboratorio de Nanoestructuras Magn\'eticas y Dispositivos, Centro At\'omico Constituyentes, B1650 Villa Maip\'u, Provincia de Buenos Aires, Argentina}

\author{A.M. Llois}
\affiliation{Instituto de Nanociencia y Nanotecnolog\'{\i}a (CNEA - CONICET), Nodo Buenos Aires, Av. Gral. Paz 1499, B1650 Villa Maip\'u, Provincia de Buenos Aires, Argentina}
\affiliation{Departamento de F\'{\i}sica de la Materia Condensada, GIyA-CNEA, Av. Gral. Paz 1499, B1650 Villa Maip\'u, Provincia de Buenos Aires, Argentina}

\author{M.H. Aguirre}
\affiliation{ Instituto de Nanociencia y Materiales de Arag\'on, CSIC, E-50018 Zaragoza, Spain.}
\affiliation{Departamento de F\'{\i}sica de la Materia Condensada, Universidad de Zaragoza, E-50009 Zaragoza, Spain.}
\affiliation{Laboratorio de Microscop\'{\i}as Avanzadas, Universidad de Zaragoza, E-50018 Zaragoza, Spain.}

\author{L.B. Steren}
\affiliation{Instituto de Nanociencia y Nanotecnolog\'{\i}a (CNEA - CONICET), Nodo Buenos Aires, Av. Gral. Paz 1499, B1650 Villa Maip\'u, Provincia de Buenos Aires, Argentina}
\affiliation{Laboratorio de Nanoestructuras Magn\'eticas y Dispositivos, Centro At\'omico Constituyentes, B1650 Villa Maip\'u, Provincia de Buenos Aires, Argentina}

\author{M.A. Barral}
\affiliation{Instituto de Nanociencia y Nanotecnolog\'{\i}a (CNEA - CONICET), Nodo Buenos Aires, Av. Gral. Paz 1499, B1650 Villa Maip\'u, Provincia de Buenos Aires, Argentina}
\affiliation{Departamento de F\'{\i}sica de la Materia Condensada, GIyA-CNEA, Av. Gral. Paz 1499, B1650 Villa Maip\'u, Provincia de Buenos Aires, Argentina}
\email{barral@tandar.cnea.gov.ar}

\begin{abstract}
$Ab$ $initio$ calculations show the presence of a strong magnetoelectric interfacial coupling in CaMnO$_3$ ultrathin film grown on a strained BaTiO$_3$ ferroelectric film. This heterostructure presents a polarization driven magnetic transition from a G-type to an A-type antiferromagnetic structure. Together with this magnetic transition we find a metallic behaviour at the interface between these two insulators, where the charge character of the carriers can be tuned from electrons to holes by switching the electric polarization direction of the FE film. Besides, the electron gas is spin-polarized while the hole-gas is not.
\end{abstract}

\pacs{0}
\maketitle

\section{Introduction}\label{intro}
\indent

During the last decades, a lot of effort has been made to design heterostructures consisting of a ferroelectric (FE) and a magnetic material, with a large interfacial magnetoelectric coupling, in order to be technologically used to control magnetism through electric fields.~\cite{Fiebig2005, Spaldin2005,Mathur2006} In the quest for functional materials, perovskites have emerged as prominent candidates due to their robust coupling among structural, orbital, charge, and spin degrees of freedom. With remarkable advancements in thin film growth techniques for perovskite oxides, precise atomic control over surfaces and interfaces has been achieved, amplifying the appeal of this material class for applications in energy, information storage, and spintronics. \\
Antiferromagnetic materials produce no demagnetization fields, are robust against magnetic field perturbations and show ultrafast dynamics with large magnetotransport responses~\cite{Wadley2016}. Therefore, particular attention should be given to antiferromagnetic manganites within magnetic perovskites. Additionally, in manganites, the exchange interactions can be regarded as a competition between ferromagnetic double-exchange (DE)~\cite{Zener1951,deGennes1960} and antiferromagnetic superexchange interactions (SE)~\cite{KRAMERS1934,Goodenough1955}. Consequently, the magnetic interaction among Mn ions is highly sensitive to lattice distortions and the local electronic and chemical environment, resulting in a diverse array of magnetic structures.\\
It is also technologically important and by now already well established that when putting together certain insulating oxides, high mobility carriers, i.e. a two-dimensional electron gas (2DEG), can emerge as a consequence of the interfacial electronic reconstruction. The most widely studied 2DEG is the one originated at SrTiO$_3$ (001)-surface~\cite{Meevasana2011} and interfaces, such as LaAlO$_3$/SrTiO$_3$ (LAO/STO)~\cite{Ohtomo2004,2007-Reyren,Brinkman2007,Stengel2011} and $\gamma$-Al$_2$O$_3$/STO.~\cite{Niu2017}  In almost all the cases, the electronic reconstruction is a consequence of either the presence of oxygen vacancies at the surface or is due to the polar discontinuity at interfaces.\cite{Mertig2020} But there might be other mechanisms yielding to the formation of a 2DEG, like the discontinuity of the charge ordering that takes place at the (001)-surface of the Peierls-like semiconductor BaBiO$_3$~\cite{Vildosola2013}, or at the BaBiO$_3$/BaPbO$_3$ interface. ~\cite{Meir2017,Dinapoli2021} Furthermore, in the last few years a lot of work has been done in order to control the interfacial electronic reconstruction with the concomitant generation of the 2DEG~\cite{2015-Mertig, Weng2021}. For instance, it was recently predicted that a 2DEG developed at the polar/non-polar LaInO$_3$/BaSnO$_3$ interface or at the ferroelectric/non-polar PbTiO$_3$/SrTiO$_3$ interface can be tuned.~\cite{Draxl2021,2023Draxl,Artacho2015}

In a previous work, we have analyzed the effect of high tensile strain and low dimensionality on the magnetic and electronic properties of CaMnO$_3$ (CMO) ultrathin films epitaxially grown on SrTiO$_3$ (STO) substrates, both, computationally and experimentally.~\cite{2020-Pedroso} From the computational simulations,  we have found that the combination of both effects yields a change in the magnetic order of CaMnO$_3$ ultrathin films, from the G-type antiferromagnetic (GAF) structure present in the bulk compound to an A-type antiferromagnetic (AAF) one. This magnetic change is also coupled to an insulator-metal transition.\\ 
In the present work, we explore if the magnetic and the insulator-metal transitions, together with the concomitant generation of a 2DEG, can be triggered and controlled by reversing the electric polarization of a ferroelectric layer in contact with the CaMnO$_3$ ultrathin film, yielding to an interfacial magnetoelectric coupling (IMEC) characterized by a magnetic reconstruction. Our goal is twofold: on one hand the pursued IMEC effect would be different and larger than the usual one,\,\cite{Tsymbal2006, Tsymbal2008,Tsymbal2010} since the change in the magnetization would be due to a change in the magnetic coupling between the Mn's moments of the ultrathin CMO film instead of coming from a variation in the magnitude of local magnetic moments.\,\cite{Burton2009, Mertig2012} On the other hand, the tunable 2DEG could be spin-polarized  due to the magnetic nature of the heterostructure.\\ 
The ferroelectric material under consideration is the tetragonal BaTiO$_3$ (BTO), which shows an electric remanent polarization in the (001) direction~\cite{1955-P_BTO}. It is worth mentioning that we choose these materials motivated by the experimental findings that show the possibility of epitaxial growth of high quality BaTiO$_3$ and CaMnO$_3$ ultrathin films on SrTiO$_3$ substrates. In our previous work, we show that CMO can be coherently grown on a (001)-SrTiO$_3$ substrate, presenting a sharp interface with negligible atomic disorder.~\cite{2020-Pedroso} Images of epitaxial BTO/STO taken by High Resolution scanning transmission electron microscopy with high angular annular dark field detector (HRSTEM-HAADF) can be found in the Supplementary Information of the present work.

\section{Computational Details} \label{comp}
\indent

We perform first-principles  calculations within  the framework of Density Functional Theory and the projector augmented wave (PAW) method~\cite{PAW}, as implemented in the Vienna \textit{ab initio} package (VASP)~\cite{VASP,PAW-VASP}. We explicitly treat 10 valence electrons for Ca (3s$^2$3p$^6$4s$^2$), 13 for Mn (3p$^6$3d$^5$4s$^2$), 10 for Ba (5s$^2$5p$^6$6s$^2$), 12 for Ti (3s$^2$3p$^6$3d$^2$4s$^2$)  and 6 for O (2s$^2$2p$^4$). The local spin density approximation (LSDA) in the parametrization of Ceperley and Alder is used.~\cite{LDA1,LDA2} As already shown, for manganese perovskites this approximation successfully predicts the observed stable magnetic phase and the structural parameters.~\cite{Picket1999, Picket2000, Nordstrom2017} Moreover, for BaTiO$_3$, the ferroelectric lattice distortion, spontaneous polarization, and lattice dynamics predicted by LDA agree well with experimental results.~\cite{Rabe2002, Rabe2003,PhysRevB.96.035143}
We include a Hubbard term with $U=5$~eV and $J=1$~eV within the Lichtenstein implementation,~\cite{Liechtenstein95} for a better treatment of the Mn $3d$-electrons in CaMnO$_3$. We have checked the robustness of our results with respect to the U and J parameters, finding that the stable magnetic structure of bulk Pnma CaMnO$_3$ corresponds to a G-type antiferromagnetic configuration, in agreement with experimental results. Moreover, the used parameters lead to a band gap of 1.60 eV, close to the experimental value of 1.55eV.~\cite{Loshkareva2004}\\

\begin{figure*}
\centering
\textit{}\includegraphics[width=2.\columnwidth]{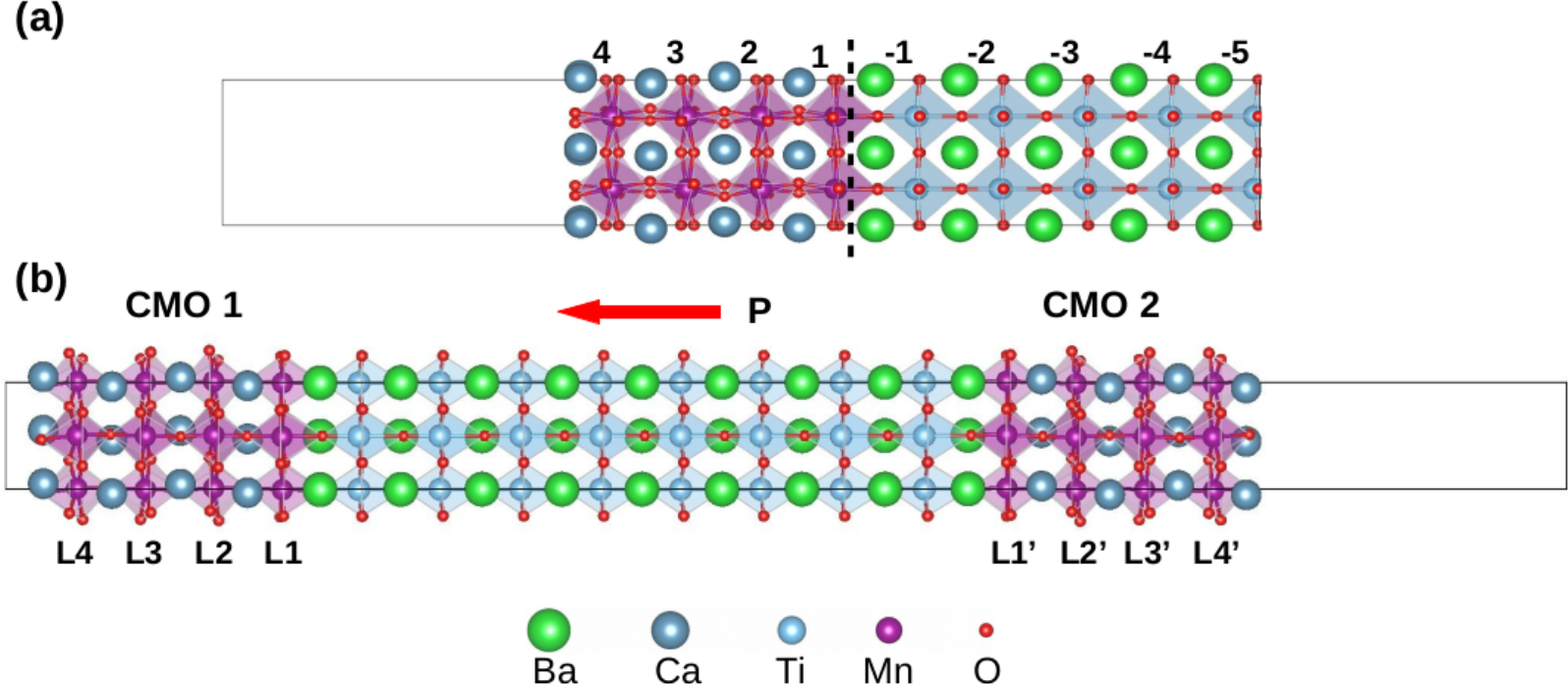}
\caption{(a) Unit cell of (CaMnO$_3$)$_4$/(BaTiO$_3$)$_5$ heterostructure. In this sketch, the polarization is pointing to the CMO film (P$\uparrow$). (b) CaMnO$_3$)$_4$/(BaTiO$_3$)$_{8.5}$/(CaMnO$_3$)$_4$ trilayer. The figures were generated using the VESTA software package \cite{VESTA}. }
\label{fig:structure}
\end{figure*}

The CaMnO$_3$/BaTiO$_3$ grown on SrTiO$_3$ slabs are modeled by a 2$\times$2 in plane supercell of the pseudo-cubic perovskite structure, with a 5 unit cells thick layer ($\sim$ 1.8nm) of BaTiO$_3$ (BTO) and a 4 unit cells thick layer of CaMnO$_3$(CMO). The MnO$_2$ termination of CMO in contact with the BaTiO$_3$ layer and the CaO termination facing the vacuum region, are considered, as depicted in Fig.~\ref{fig:structure}(a). 
The SrTiO$_3$ (STO) substrate is not explicitly treated, but is implicitly included by constraining the in-plane lattice constant of the whole structure to the optimized bulk cell parameter of STO ($a_{STO}$=3.87~\AA). The considered BTO thickness was demonstrated to be sufficient for the ferroelectric instability to develop.~\cite{Rabe2006, Gerra2006, Tsymbal2006, Mertig2008} At the same time, it is thin enough to be fully strained when grown on the substrate (See SI).  

 We relax the internal coordinates of all atoms within the CMO film and the two BTO layers adjacent to the CMO. The remaining atoms in the BTO film are fixed at their bulklike and previously optimized positions in the strained bulk calculations, in order to fix its ferroelectric polarization, which was determined from first principles using the Berry-phase formalism.~\cite{Vanderbilt1993} A vacuum spacer of 17.5\,\AA \, along the z-axis was used in order to avoid the self-interaction through the periodic boundary conditions. In addition, since the BTO layer has a net dipole moment oriented along the (001)-direction, the dipole correction, as implemented in VASP, was used. We call the ferroelectric polarization \textbf{P} ‘up’ (P$\uparrow$) when it is pointing to the interface and ‘down’ (P$\downarrow$) when it is pointing away from the CMO/BTO interface. \\
Similar to our previous work~\cite{2020-Pedroso}, we consider the most relevant antiferromagnetic (AF)  orders that might be shown by the Mn atoms, namely AAF (ferromagnetic (FM) ordered (001) planes that align antiparallel with neighboring planes along the [001] direction), CAF (AF alignment within the (001) planes and FM between adjacent planes) and GAF (AF coupling between first nearest neighbours). It is worth mentioning that the noncollinearity of the Mn magnetic structure is known to be quite minimal and gives rise to a small magnetic moment of 0.04$\mu_B$\,\cite{Spaldin2011,Bibes2019} and, therefore we approximate the magnetic configuration with a collinear model. \\
All the DFT calculations are performed using a 500~eV energy cutoff in the plane waves basis. We use a 6$\times$6$\times$1 Monkhorst-Pack $k$-point grid centered at the $\Gamma$-point to evaluate integrals within the Brillouin zone (BZ). 
The structural relaxations are performed until the forces on each ion are less than 0.01~eV/\AA. \\

We also address the electronic and magnetic properties of the interface between strained CaMnO$_3$ and  BaTiO$_3$ using an alternative structural model consisting of a (CaMnO$_3$)$_4$/(BaTiO$_3$)$_{8.5}$/(CaMnO$_3$)$_4$ (CMO1 /BTO /CMO2) trilayer. The non-stoichiometry of the BTO film is necessary in order to obtain the BaO-termination of both interfaces.  The in plane unit cell is a  $\sqrt2\times\sqrt2$  supercell of the pseudo-cubic perovskite structure containing two transition metals (Mn/Ti) per plane and a vacuum spacer along the z-axis of 18.8\,\AA, as depicted in Fig.~\ref{fig:structure}(b). All the computation parameters (plane wave's energy cutoff, k-points grid, ionic relaxation and dipole correction) are set to guarantee the convergence of our results. Within this structural model, the CMO1/BTO interface represents the previous P$\uparrow$ regime while the BTO/CMO2 corresponds to the P$\downarrow$ one and, therefore, we can simulate both polarization regimes simultaneously. 

\section{Results and discussion}\label{results}
\indent
As already known for more than three decades, bulk crystalline BTO shows a cubic symmetry and is paraelectric above T$_c$=393~K, becomes tetragonal and ferroelectric below T$_c$, and goes through additional transitions to orthorhombic at 278~K and rhombohedral at 183 K.~\cite{Kwei1993} In its tetragonal phase, BaTiO$_3$ shows an electric remanent polarization P$_{BTO} \simeq 26 \mu$C/cm$^2$ in the (001) direction~\cite{1955-P_BTO}.\\
As a first step, we calculate the polarization value of tetragonal BaTiO$_3$ when it is epitaxially grown on top of a cubic (001) SrTiO$_3$ substrate, whose lattice constant was previously optimized within the LDA. Under compressive epitaxial strain, its polarization increases with increasing tetragonality and was predicted to be enhanced for large compressive misfit, as the one imposed by a coherent epitaxial growth on perovskite substrates having a smaller lattice constant, such as SrTiO$_3$.~\cite{Rabe2002} Under this compressive misfit strain of $\simeq -2.2\%$ imposed by the substrate, the spontaneous polarization is enhanced by 34$\%$ (P$_{BTO/STO}$ = 37.4 $\mu$C/cm$^2$) compared to its value in the unconstrained bulk tetragonal phase (P$_{BTO}$ = 27.9 $\mu$C/cm$^2$).
In Table~\ref{tab:BTO_P} we show the computed P values for different in-plane ($a$) parameters, corresponding to the experimental value, the LDA relaxed BaTiO$_3$ and SrTiO$_3$ in-plane lattice constants. In all the cases, the out-of-plane ($c$) lattice parameters were optimized.  As expected, we find an enhancement of the ferroelectric polarization due to the compressive strain, with the corresponding elongation of the $c$-parameter, in agreement with previous reported calculations.~\cite{Rabe2002,Rabe2003}

\begin{table}[ht!]
    \centering
    \caption{BTO polarization values along the (001)-direction, calculated for different in-plane ($a$) and out-of-plane ($c$) lattice parameters. }
    \begin{tabular}{|c|c|c|c|}
    \hline
    &$a$ (\AA)     &  $c$ (\AA) & $P_z$ ($\mu C/cm^2$) \\
    \hline
    \hline
    experimental BTO~\cite{Kwei1993} & 3.991 & 4.035   & 26.0\\
    bulk  BTO & 3.942 & 4.004   & 27.9\\
    strained bulk BTO  & 3.867 & 4.100   & 37.4\\
    \hline 
    \hline
     \end{tabular}
    \label{tab:BTO_P}
\end{table}

To study the influence of the FE polarization switching in the magnetic properties of the CMO film, we consider  a (CaMnO$_3$)$_4$/(BaTiO$_3$)$_5$ bilayer, as the one depicted in Fig.~\ref{fig:structure}(a). We take into account three polarization regimes, that is P$\uparrow$, P$\downarrow$ and P$=0$, where the last one corresponds to the centrosymmetric structure of BaTiO$_3$. In Table~\ref{tab:enebilayer}, the total energies referred to the GAF-magnetic structure within each P-regime are informed. As it can be seen, the GAF-type is the most stable magnetic structure for P$=0$ and P$\downarrow$ polarizations, while the AAF-type is the ground state in the case of P$\uparrow$. 

\begin{table}[ht!]
\centering
\caption{Energy with respect to the corresponding GAF-magnetic configuration (in meV/f.u.) for the different orientations of the BaTiO$_3$ polarization in the (CMO)$_4$/(BTO)$_5$ heterostructure. The energies corresponding to zero polarization is also shown, for comparison. }
\label{tab-CMO-bulk}
\begin{tabular}{|c|c|c|c|}
\hline
Mag.Conf. & \textbf{P$\uparrow$} & \textbf{P$\downarrow$} &\textbf{P = 0} \\
\hline
\hline
GAF	 &   0.0   &   0.0 &   0.0   \\
AAF  & -18.0 &    8.3 &  10.7  \\
CAF  &  11.6 &    9.8 &  12.1 \\
FM   &  -0.9 &   22.4 &  27.5  \\
\hline	
\hline
\end{tabular}
\label{tab:enebilayer}
\end{table}

\begin{figure*}
\centering
\includegraphics[width=1.\columnwidth]{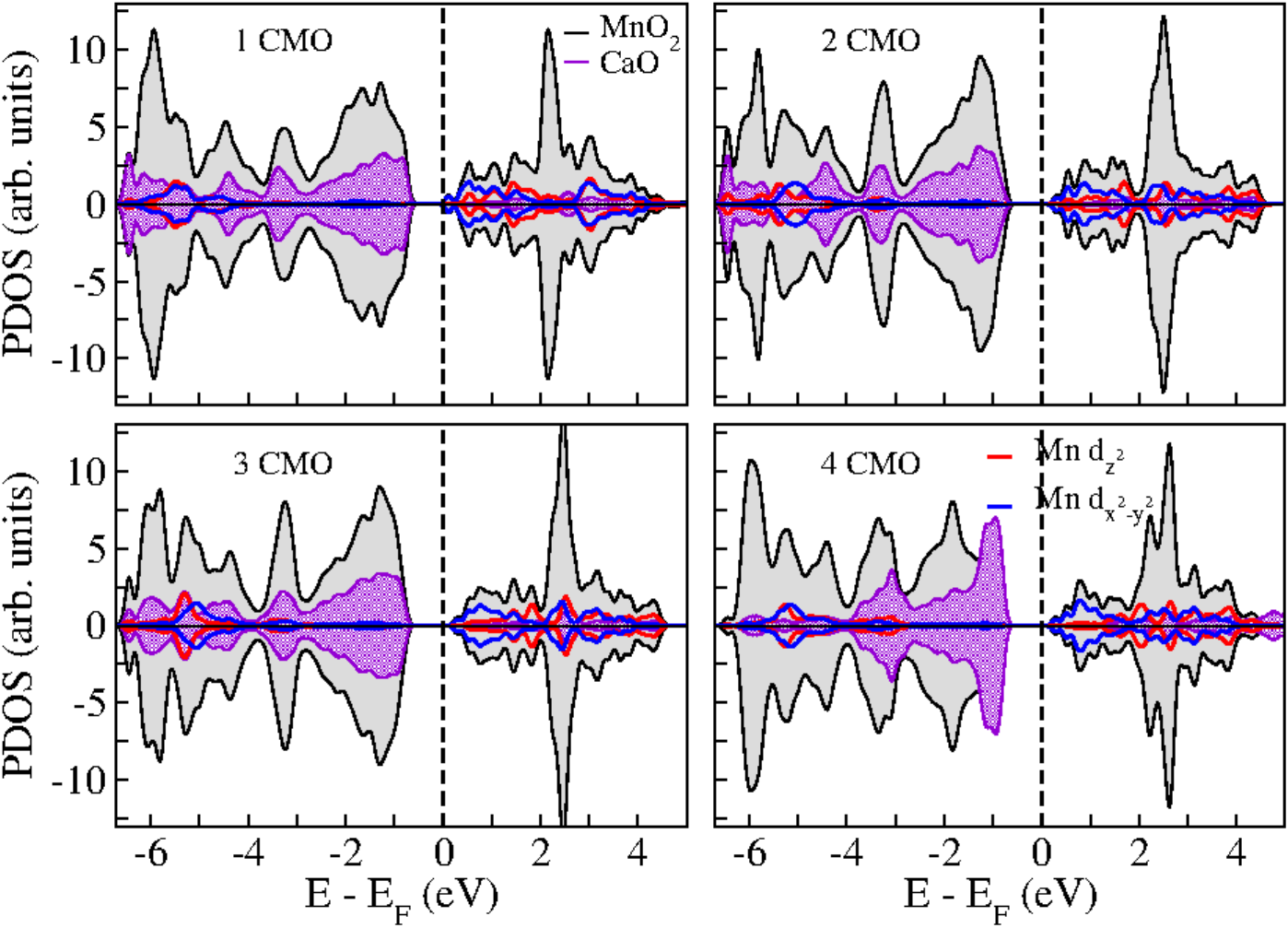}
\includegraphics[width=1.\columnwidth]{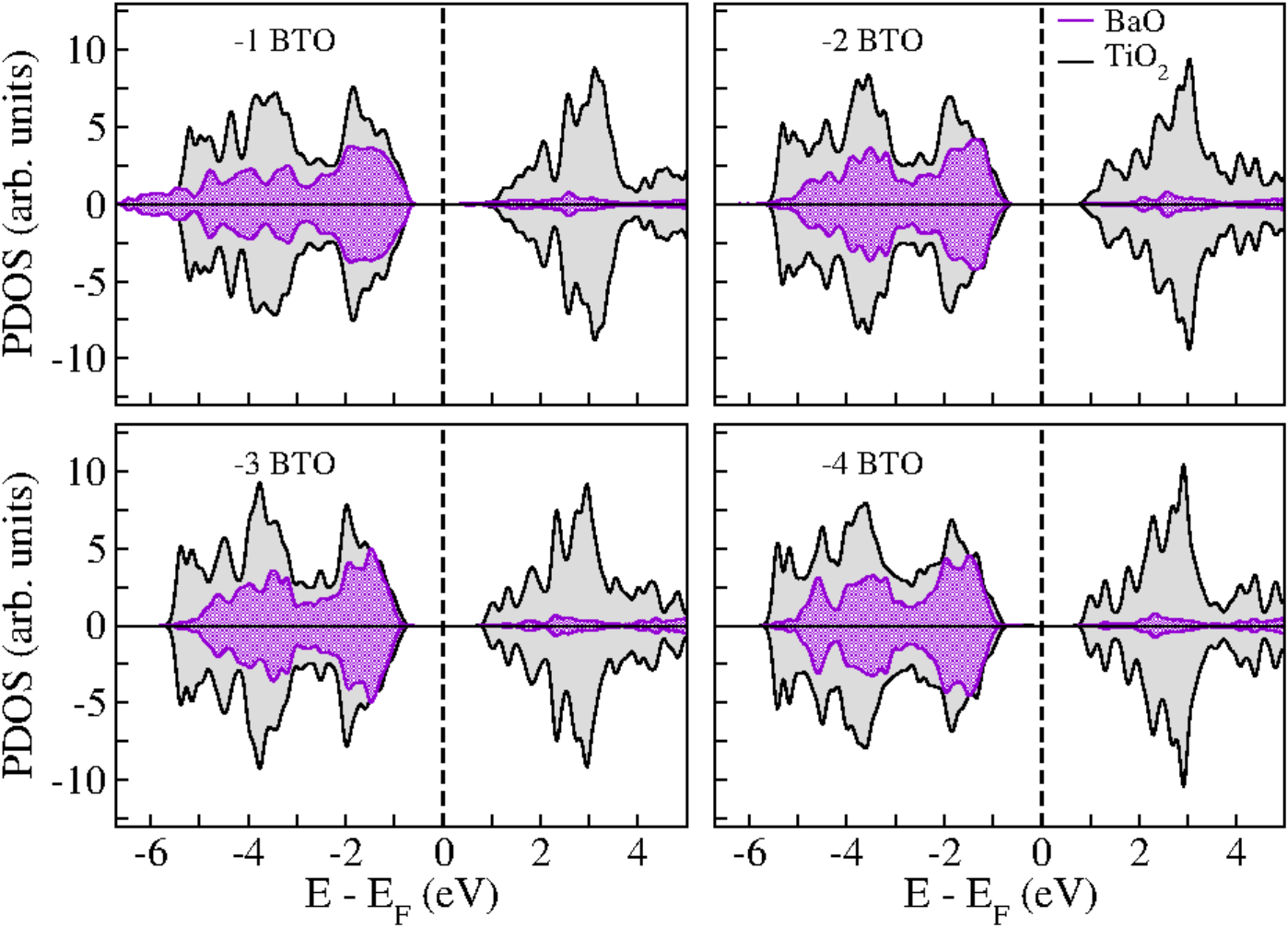}
\caption{(Color online) Layer resolved PDOS of CMO (left) and BTO (right) layers (excluding the BTO-vacuum interface), corresponding to P=0. In blue (red) the Mn d$_{x^2-y^2}$ (d$_{z^2}$) character is shown.}
\label{fig:PDOS-P0}
\end{figure*}

\begin{figure*}[ht!]
\centering
\includegraphics[width=1.\columnwidth]{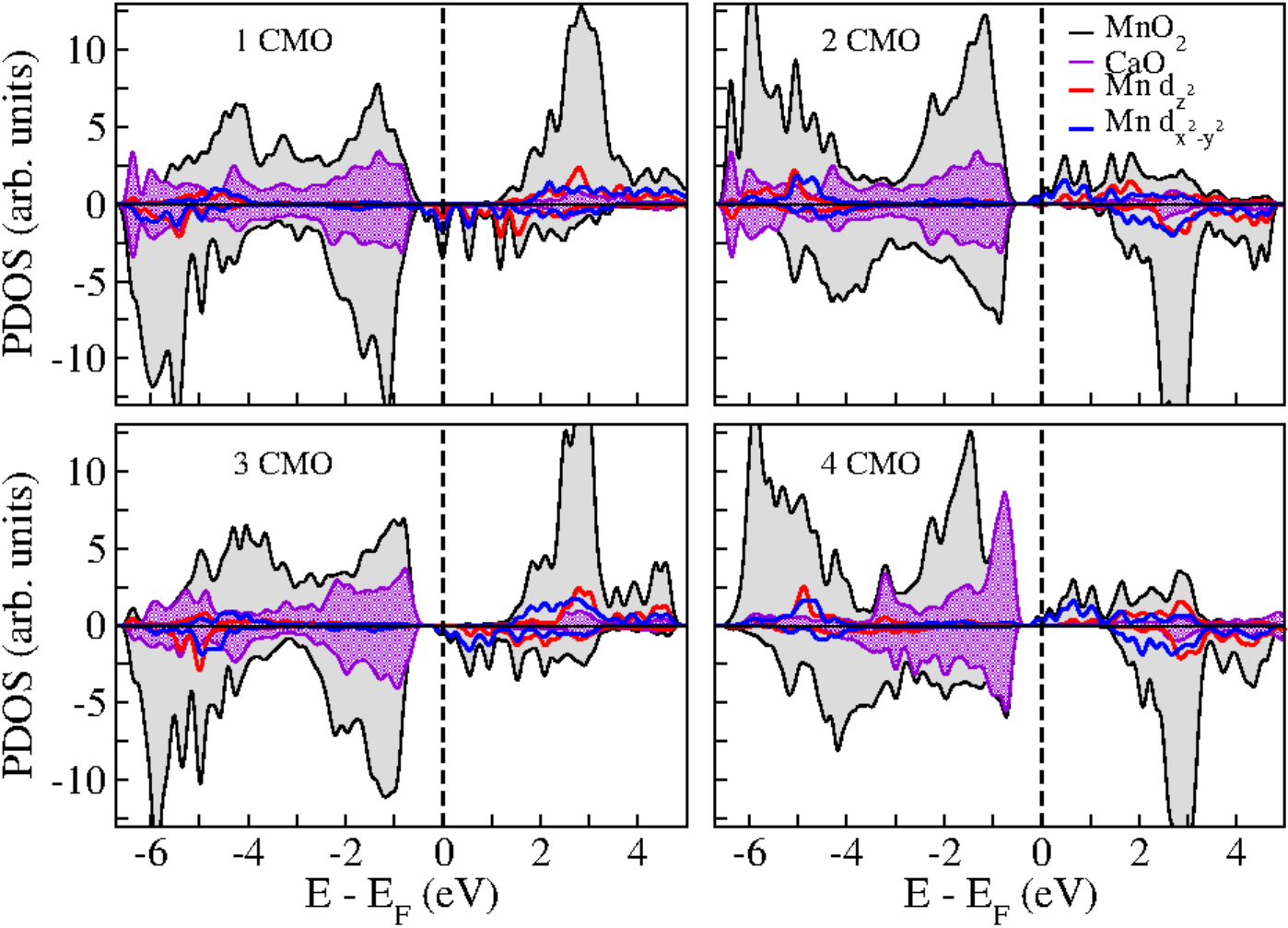}
\includegraphics[width=1.\columnwidth]{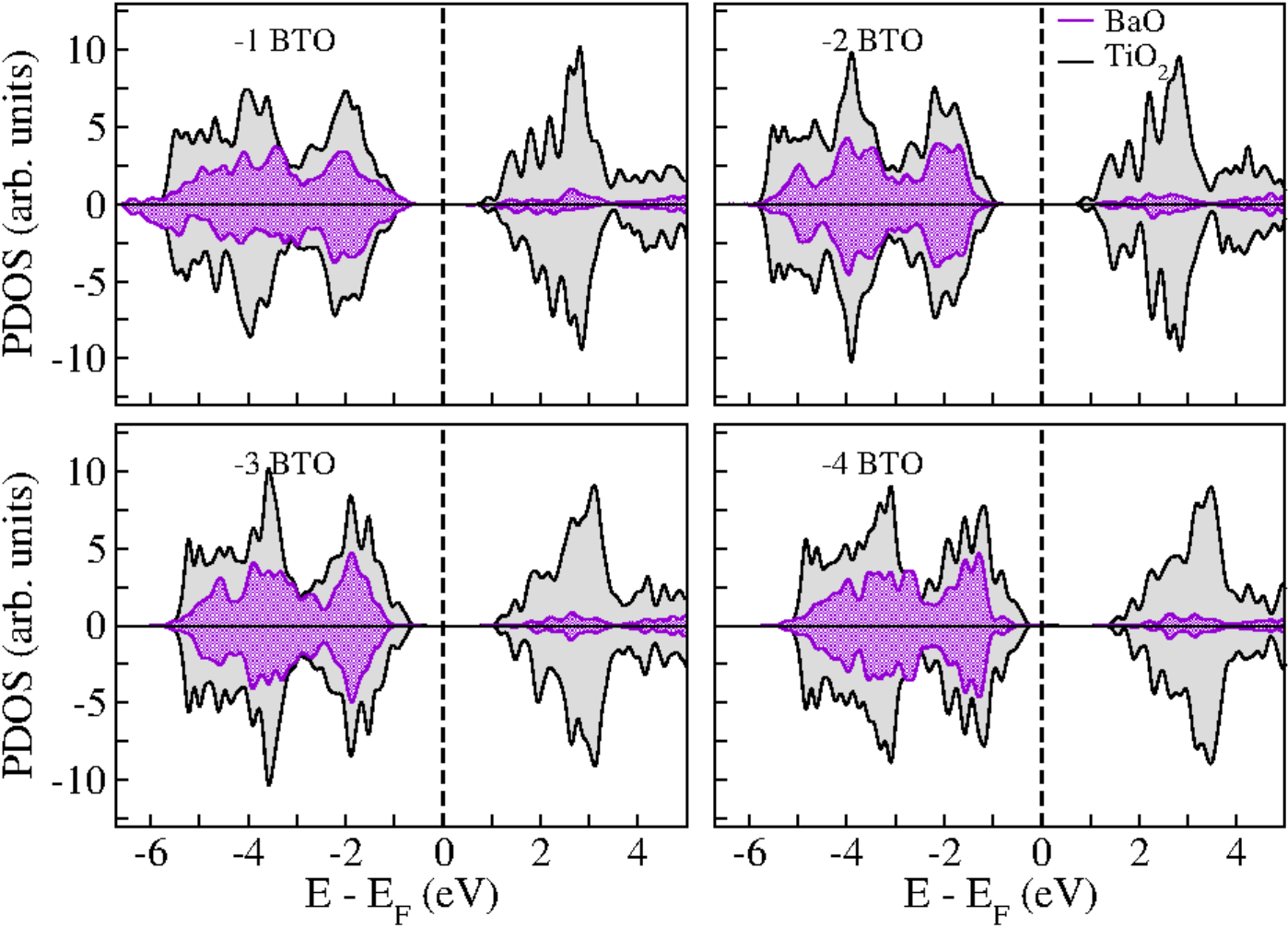}\\
\includegraphics[width=1.\columnwidth]{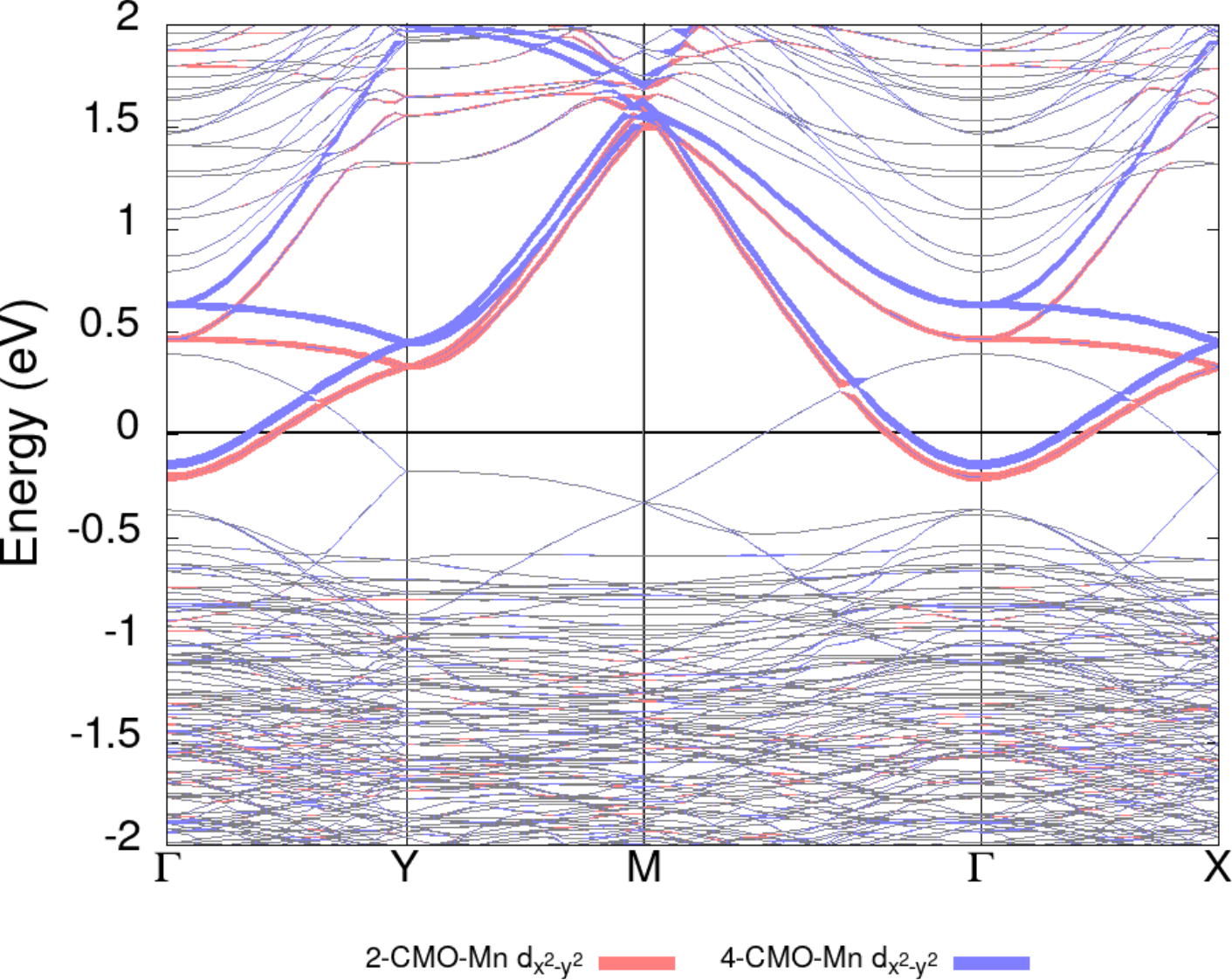}
\includegraphics[width=1.\columnwidth]{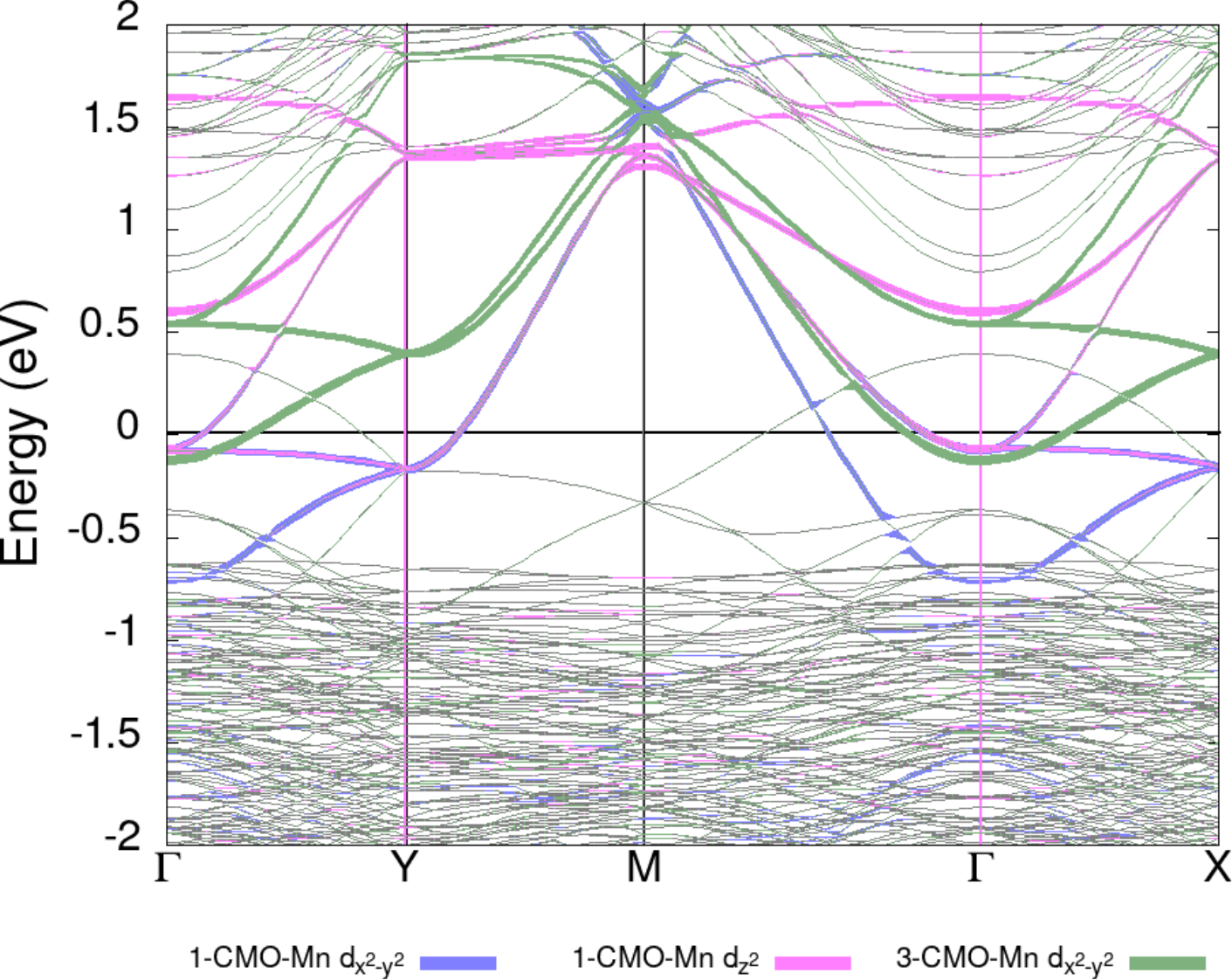}
\caption{(Color online) Upper panels: layer resolved PDOS of CMO (left) and BTO (right) layers (excluding the BTO-vacuum interface) for the case of polarization pointing to the CMO interface  (P$\uparrow$). Lower panels: spin up (left) and spin down (right) bandstructure of the heterostructure. The different character of the Mn's $e_g$ orbitals coming from each CMO layer is highlighted.}
\label{fig:Pup}
\end{figure*}

\begin{figure*}[ht!]
\centering
\includegraphics[width=1.\columnwidth]{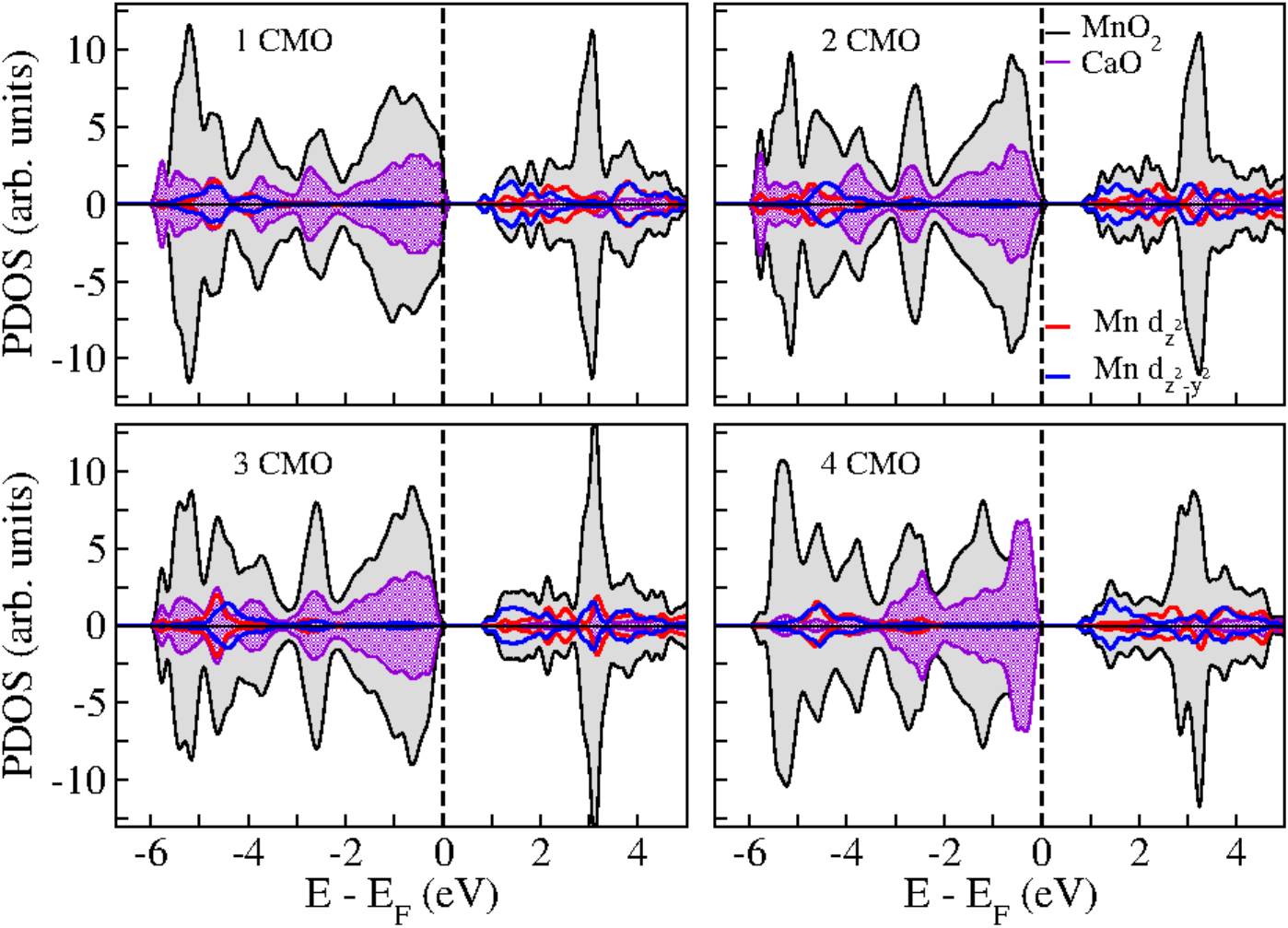}
\includegraphics[width=1.\columnwidth]{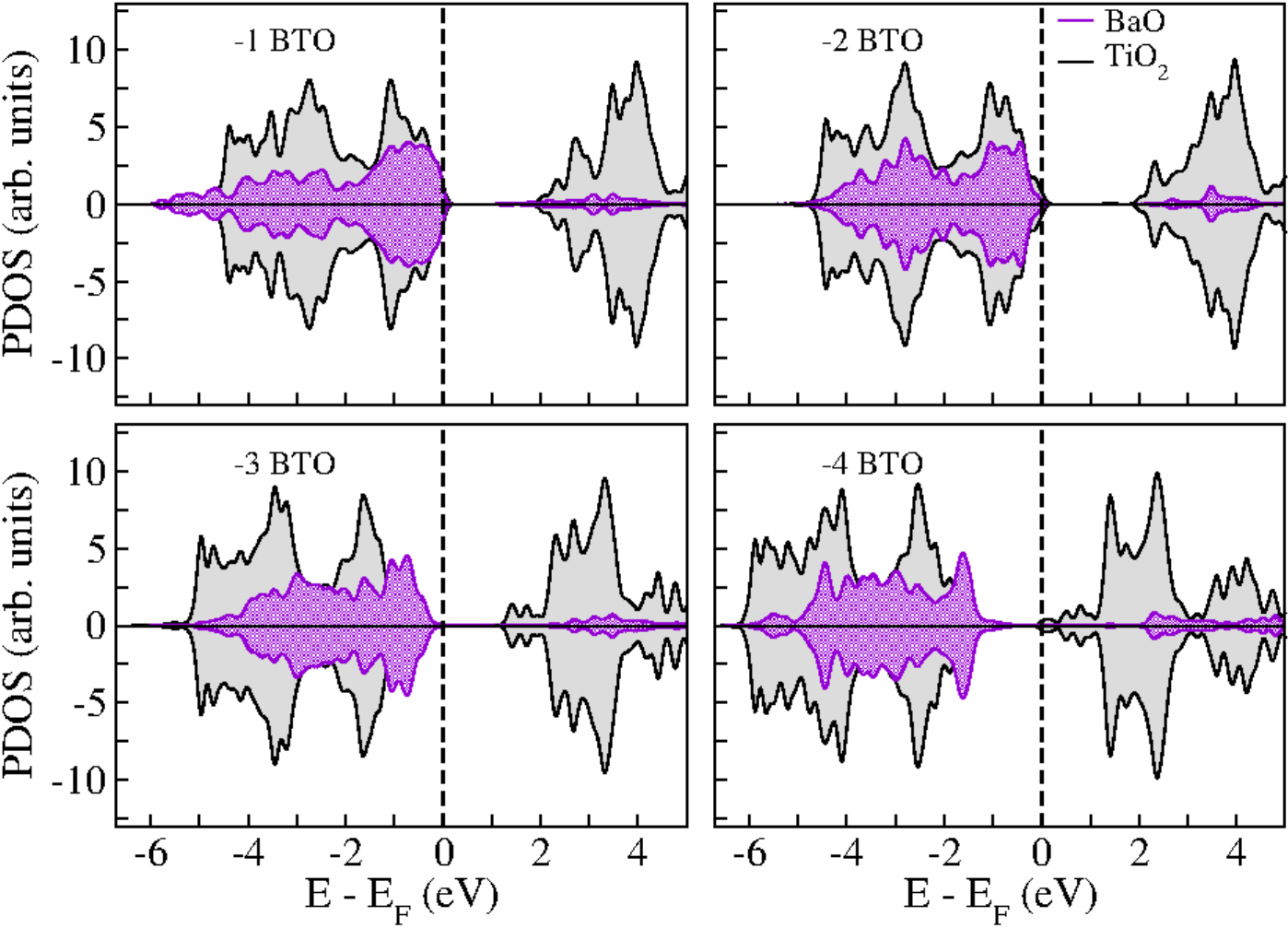}\\
\includegraphics[width=1.\columnwidth]{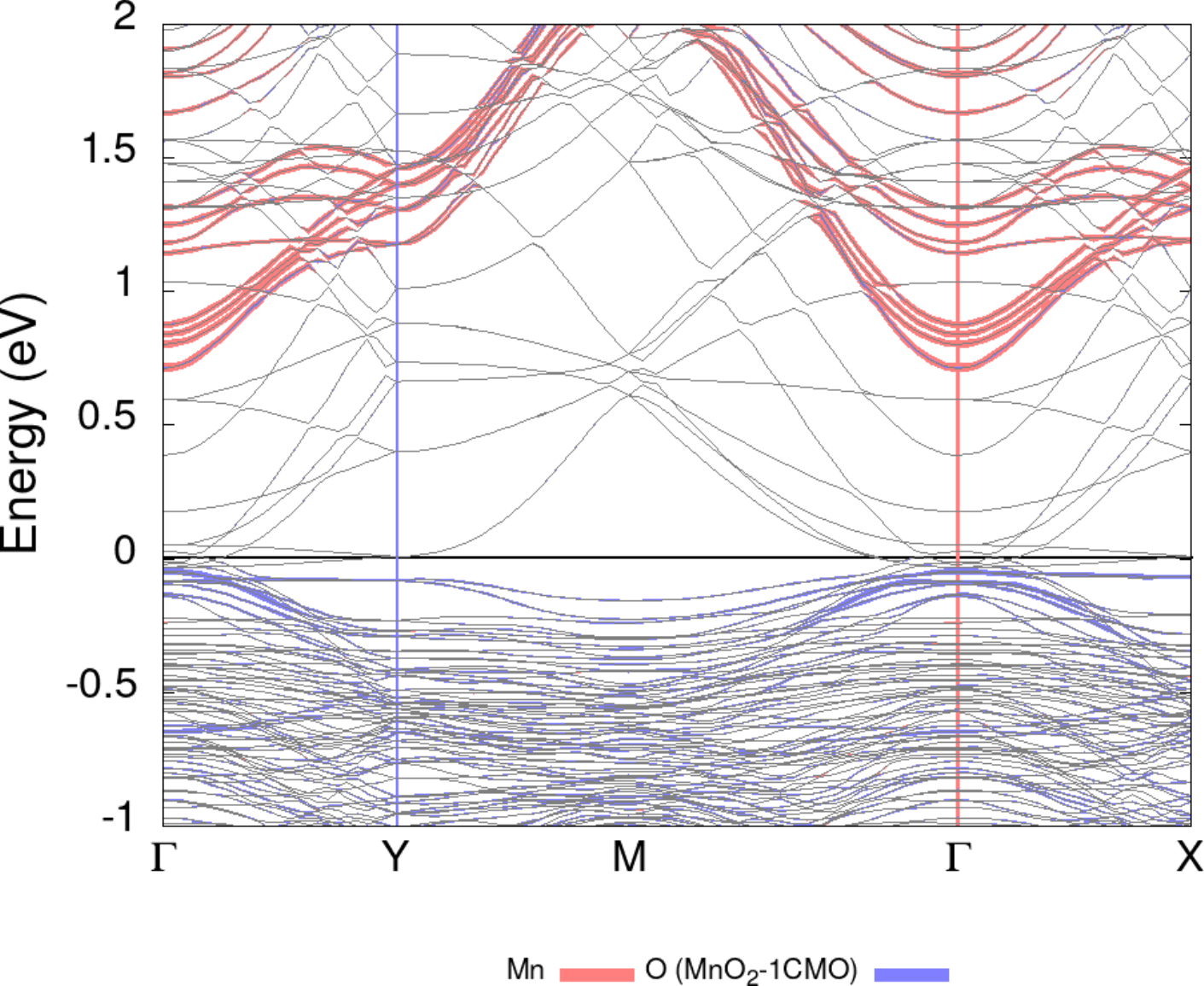}
\includegraphics[width=1.\columnwidth]{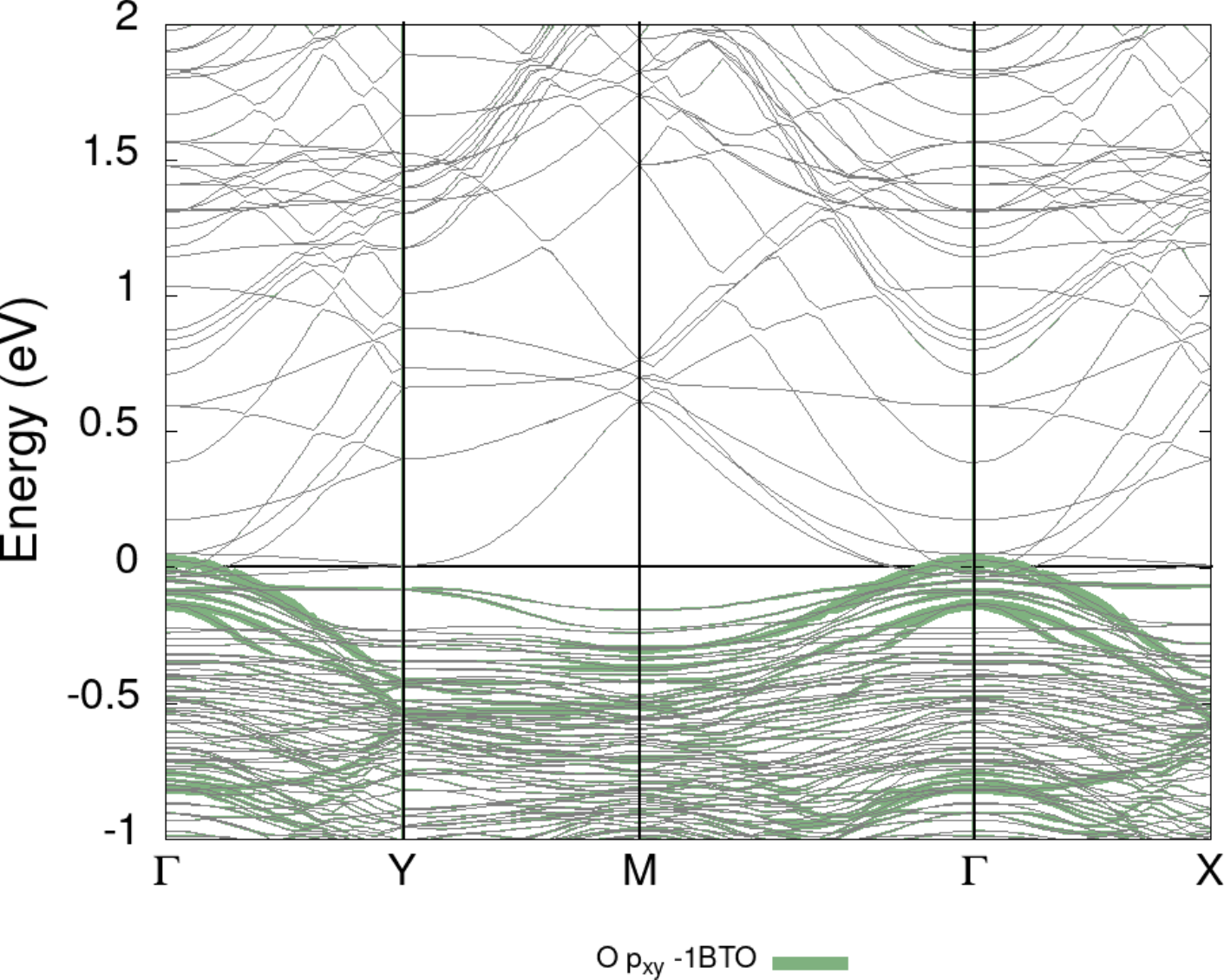}
\caption{(Color online) Upper panels: layer resolved PDOS of CMO (left) and BTO (right) layers (excluding the BTO-vacuum interface) for the case of polarization pointing away from the CMO interface  (P$\downarrow$). Lower panels: spin(up+down) bandstructure of the heterostructure. Projections onto the Mn-$e_g$ and interfacial O(CMO)- orbitals (left) and onto interfacial O-$p_x+p_y$ (BTO) (right) are highlighted. Ti-bands are above 2.5 eV}
\label{fig:Pdn}
\end{figure*}

To understand the origin of this ferroelectric-driven magnetic transition, we analyze the layer-resolved partial density of states (PDOS) for the ground states of the P$\uparrow$ and P$\downarrow$ regimes and compare them to the P$=0$ case (See Figs.~\ref{fig:PDOS-P0},~\ref{fig:Pup} and \ref{fig:Pdn}). For the nonpolarized centrosymmetric BTO structure, we find that the whole heterostructure remains insulating. It is interesting to note for our future analysis that, due to the tensile strain in the CMO film, the degeneracy of the two empty Mn's $e_g$ orbitals is lifted and the bottom of the conduction band has a d$_{x^2-y^2}$ character, as can be seen in the left panel of Fig.~\ref{fig:PDOS-P0}.\\

When the polarization points to the CMO film an electronic reconstruction emerges in order to screen the positive polarization charge, as shown in the left upper panel of Fig.~\ref{fig:Pup}. It can be seen from this figure that charge transferred from the BTO to the CMO film begins to fill the empty Mn-$d$ conduction bands and this effect continues through all the 4 CMO layers. Due to the broken degeneracy of the $e_g$ orbitals, the ones which begin to be partially occupied are the d$_{x^2-y^2}$ ones. As known, the magnetic interaction between the Mn atoms is determined by the competition between AFM superexchange via the Mn $t_{2g}$ electrons and FM double exchange via the Mn $e_g$ electrons. When the d$_{x^2-y^2}$ orbitals are partially occupied the double exchange in the MnO$_2$ planes strengthens the FM ordering while superexchange stabilizes the AFM one between the planes, leading to the AAF magnetic ground state. Coupled to this polarization driven magnetic transition, we find an insulator to metal transition characterized by an extended two dimensional electron gas (2DEG) which is mainly localized at the interface. A rough estimation of this charge transfer can be calculated from the computed polarization value of 37.4$\mu C/cm^2$, which gives a net charge of around 1.4$e^-$ when taking into account our surface unit cell. This value is consistent with the one obtained when integrating the density of states corresponding to the CMO conduction bands, namely  $\simeq 1.03 e^-$, which is the charge inside the atomic spheres given by the used pseudopotentials. This charge is spin polarized, with a net magnetic moment of 0.2 $\mu_B$,  due to the uncompensated magnetic moments between the CMO layers (See Table~\ref{tab:mmbilayer}). This feature can also be observed in the bottom panels of Fig.~\ref{fig:Pup}, where we show the spin up (left) and spin down (right) bandstructure projected onto the Mn-d$_{x^2-y^2}$ bands of all the layers. For the interfacial Mn atoms we also plot the d$_{z^2}$ character of the bands, which are slightly occupied. It can be noted that the spin-down Mn-d$_{x^2-y^2}$ band corresponding to the interfacial 1-CMO layer is shifted with respect to the spin-up Mn-d$_{x^2-y^2}$ of the 2-CMO one, which gives rise to the uncompensated magnetic moment of the whole CMO film. As expected, the electronic reconstruction induced by the FE polarization decreases when going far from the interface. The interfacial band that crosses the Fermi level (highlighted in blue in the right bottom panel of Fig.~\ref{fig:Pup}) presents a slightly anisotropic dispersion around the $\Gamma$-point with estimated effective masses of 0.48$m_e$ and 0.37$m_e$, for the $\Gamma$-X  and the $\Gamma$-M directions, respectively. These values are of the same order as the ones reported for the 2DEG generated at the SrTiO$_3$/LaAlO$_3$ interfaces.~\cite{Mertig2020}\\

\begin{table}[h!]
\centering
\caption{Mn magnetic moments (in $\mu _B$) within each MnO$_2$ layer, from 1 to 4, as labeled in Fig.\ref{fig:structure}(a). }
\label{tab-CMO-bulk}
\begin{tabular}{|c|c|c|c|c|c|}
\hline
&		& \textbf{1} &	\textbf{2} & \textbf{3} & \textbf{4}\\
\hline \hline
\textbf{P=0}   &GAF	& $\pm$ 2.74 & $\mp$ 2.78 & $\pm$ 2.78 &  $\mp$ 2.79 \\
\hline	
\textbf{P$\uparrow$} &AAF &  -3.05     &   2.85     &  -2.85     & 2.85 \\        
\hline	
\textbf{P$\downarrow$}&GAF	& $\pm$ 2.74 & $\mp$ 2.78 & $\pm$ 2.78 &  $\mp$ 2.79 \\
 \hline	\hline
\end{tabular}
\label{tab:mmbilayer}
\end{table}

When the polarization is pointing away from the CMO-film (P$\downarrow$) we find a slight electron's depletion at the first two layers of the BTO film (i.e. -1 BTO and -2 BTO), which gives rise to a two dimensional hole gas (2DHG) located within these two layers (See upper panels of Fig.~\ref{fig:Pdn}). For this polarization direction, the CMO film remains insulating with the three electrons of each Mn occupying the $t_{2g}$ orbitals thus leading to the GAF magnetic ordering. The local magnetic moments of each MnO$_2$ plane are shown in Table~\ref{tab:mmbilayer}, where it can be seen that the layer by layer magnetic moments coincide with the ones obtained in the P=0 regime. In the lower-left panel of Fig.~\ref{fig:Pdn} the character projected bandstructure  onto the Mn-$e_g$ orbitals (red) and onto the O orbitals of the interfacial CMO layer (blue) are depicted, confirming the insulating behavior of the CMO film. From the lower-right panel of the same figure, it can be seen that the generated 2DHG comes from the interfacial oxygens of the BTO film, and it presents a non spin-polarized p$_x$+p$_y$ character. As the magnetic order of this P-regime corresponds to the G-AFM type, the total magnetic moment within each layer vanishes and, therefore, we plot only one spin-projection of the bandstructure. \\
It is important to note that in the bandstructures of Figs.~\ref{fig:Pup} and~\ref{fig:Pdn} there are other grey-coloured bands that are crossing the Fermi level (not highlighted) which are spatially localized at the BTO-vacuum interface. This fact is obviously a consequence of the charge neutrality of the whole system.

\begin{figure*}
\includegraphics[width=2.\columnwidth]{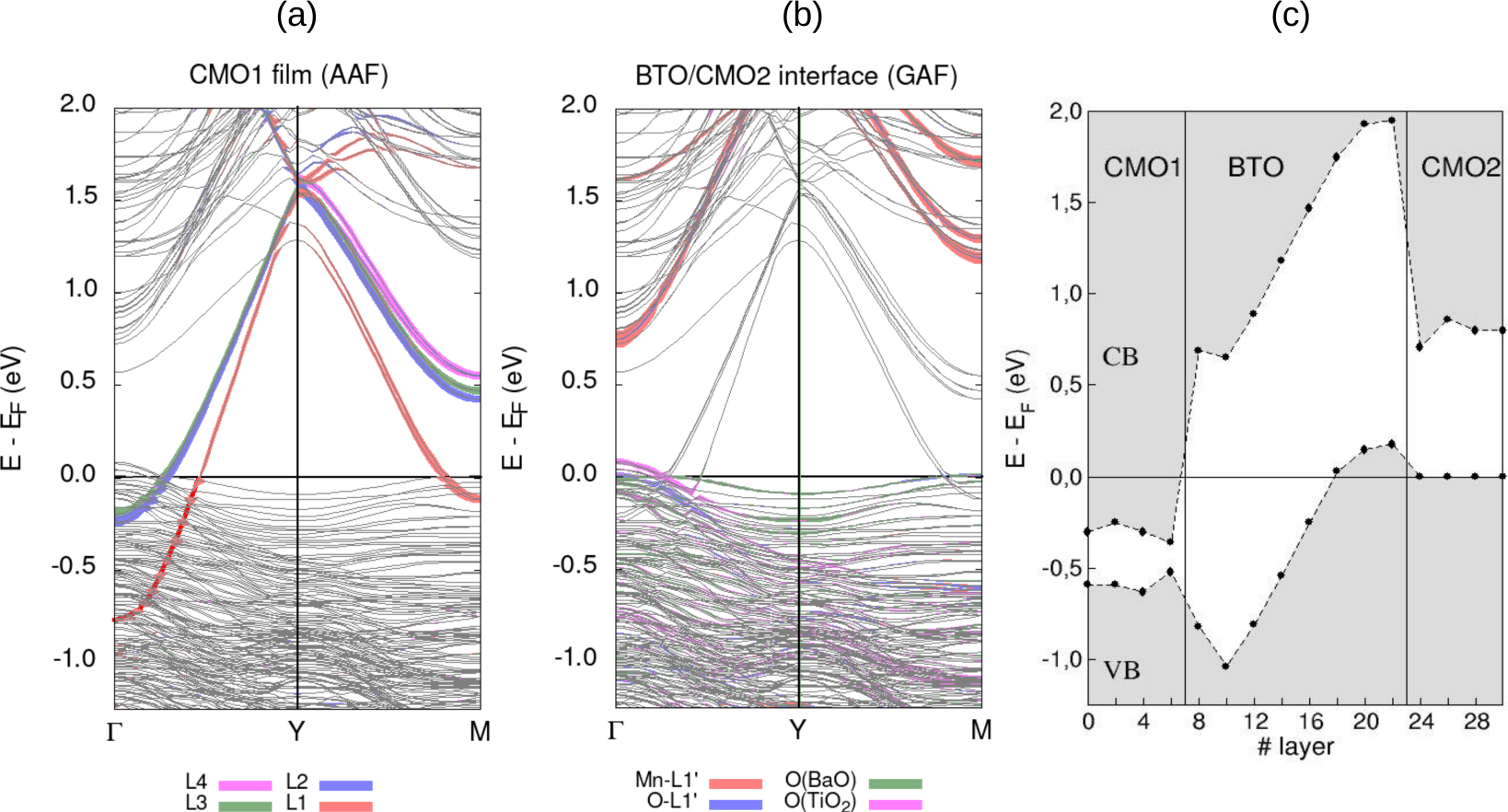}
\caption{(Color online) Spin up + spin down bandstructure of the trilayer corresponding to its ground state AAF-GAF. (a) projections onto the Mn-d$_{xy}$ orbitals for the different MnO$_2$ layers within the CMO1-film are highlighted (See Fig.~\ref{fig:structure}(b)).  (b) projections onto the p$_x$+p$_y$-orbitals of O atoms within both CMO2 and BTO interfacial layers. Mn-$d$ orbitals are above 0.5eV, as shown in the red-highlighed curve. (c) Calculated CB and VB profiles for the [001] direction of the trilayer. Note that layer $\#1$ in this figure corresponds to layer named L4 in Fig.~\ref{fig:structure}(b)} 
\label{fig:trilayer_bands}
\end{figure*}

To strengthen the robustness of our findings at the CMO/BTO interface we also analyze the electronic properties of a (CMO)$_4$/(BTO)$_{8.5}$/(CMO)$_4$ trilayer, where the two non-zero polarization regimes can be simultaneously simulated (See Fig.~\ref{fig:structure}(b)). Taking into account our previous results, the CAF and the FM magnetic configurations of both the CMO1 and the CMO2 films are not considered. Therefore, we calculate only two different magnetic structures of the two films, obtaining the following four combinations: AAF-AAF, AAF-GAF, GAF-AAF and GAF-GAF. As already mentioned in Section~\ref{comp}, within this structural model, the CMO1/BTO interface represents the previous P$\uparrow$ regime while the BTO/CMO2 corresponds to the P$\downarrow$ one. It is straightforward to infer, from our total energy results presented in Table~\ref{tab:trilayer}, that the ground state magnetic configuration is the AAF-GAF one, in agreement with the bilayer's results.\\
Further analysis can be obtained from the electronic properties of the system close to the Fermi level. In Fig.~\ref{fig:trilayer_bands} we show the corresponding bandstructure highlighting the projection of the bands onto the d$_{xy}$-orbitals of the Mn atoms of the CMO1-film (a) and onto the oxygen atoms located close to the BTO-CMO2 interface (b). From these two figures the existence of the two dimensional electron and hole gases at the two interfaces is confirmed. In the case of negative carriers, this 2DEG is located at the CMO1-BTO interface that corresponds to the P$\uparrow$-regime and it is generated when the d$_{xy}$-orbitals of the Mn atoms begin to be partially filled. It is important to note that the  unit cell of the trilayer is rotated 45º with respect to the bilayer one and, therefore, the d$_{xy}$ orbitals of the first correspond to the d$_{x^2-y^2}$ orbitals of the second system. 
At the BTO-CMO2 interface (P$\downarrow$-regime) the charge carriers have positive character and come from the p$_x$+p$_y$ orbitals of the interfacial oxygen atoms that start to be partially empty. The effective mass of the 2DHG is isotropic and its value is 0.96$m_e$, which is also of the same order as the ones reported for the 2DHG generated at the SrTiO$_3$/LaAlO$_3$ interfaces.~\cite{Mertig2020} To summarize our results for the trilayer system, in Fig.~\ref{fig:trilayer_bands}(c) we show the band bending of the heterostructure. As it can be observed from this figure, where we plot, both, the top of the valence band and the bottom of the conduction band along the growth direction of the film, the 2DEG is spread within the whole CMO1 film, while the 2DHG is localized at the BTO side of the BTO/CMO2 interface.

\begin{table}
\centering
\caption{Energy with respect to the equilibrium magnetic configuration, $\Delta E$, in meV/f.u., for both CMO1 and CMO2 layers (See Fig.\ref{fig:structure}(b)). }
\label{tab-CMO-bulk}
\begin{tabular}{|c|c|}
\hline
CMO1-CMO2 magnetic configuration &  $\Delta E$  \\
\hline
\hline
AAF - AAF    &   4.4  \\
AAF - GAF    &   0.0    \\
GAF - AAF    &  13.1    \\
GAF - GAF    &   9.4   \\
\hline		
\hline
\end{tabular}
\label{tab:trilayer}
\end{table}

\section{Conclusions}
In this work we study, by means of $ab$ $initio$ calculations within the Density Functional Theory, the electronic and magnetic properties of a CaMnO$_3$ ultrathin film grown on a strained BaTiO$_3$ ferroelectric film. 
We claim that this heterostructure presents a strong magnetoelectric interfacial coupling, characterized by a polarization driven magnetic transition from a G-type to an A-type antiferromagnetic configuration. Furthermore, coupled to this MEC we find a metallic behaviour at the interface between these two insulators, where the charge character of the carriers can be tuned from electrons to holes by switching the electric polarization direction of the FE film. When this polarization points towards the CMO-film, we find a spin-polarized 2DEG hosted by the Mn-3d$_{x^2-y^2}$ bands spatially localized within all the CMO layers,  with a slight anisotropic dispersion around the $\Gamma$-point with estimated effective masses of 0.48$m_e$ and 0.37$m_e$, for the $\Gamma$-X  and the $\Gamma$-M directions, respectively. On the other hand, when the FE polarization points away from the CMO interface, the charge carriers are of the holes type, and this 2DHG is non spin-polarized with an O-p$_x$+p$_y$ character, spatially localized within the first interfacial BTO layer.\\
Summarizing, our results show that the magnetic ordering of the CMO and, consequently, the exchange interactions of the Mn atoms can be controlled by the ferroelectric polarization direction of BTO in this strained heterostructure. The polarization of FE film can be reversed using an external electric field, allowing us to manipulate the charge transfer at the interface and change the occupancy of the Mn $e_g$ orbitals, thereby modifying the magnetic ordering of the ultrathin CMO film.\\
Overall, the proposed heterostructure offers unique potential opportunities to be considered as a fundamental part of future technological applications in spintronic devices. We hope that our findings stimulate further experimental studies towards miniaturization and interface engineering where the interfacial magnetoelectric coupling triggers and controls the physical properties.\\ 

\section*{Acknowledgments}
This work was partially supported by PICT-2016-0867 and PICT-2019-2128 of the ANPCyT, Argentina, and by H2020-MSCA-RISE-2016 SPICOLOST Project Nº 734187. We also acknowledge the financial support of European Commission through Marie Skłodowska-Curie Actions H2020 RISE with the projects MELON (grant no. 872631) and ULTIMATE-I (grant no. 101007825). Authors would like to acknowledge the access of equipment of "Servicio General de Apoyo a la Investigaci\'on (SAI), Universidad de Zaragoza".

\bibliography{biblio.bib}

\begin{thebibliography}{51}
\expandafter\ifx\csname natexlab\endcsname\relax\def\natexlab#1{#1}\fi
\expandafter\ifx\csname bibnamefont\endcsname\relax
  \def\bibnamefont#1{#1}\fi
\expandafter\ifx\csname bibfnamefont\endcsname\relax
  \def\bibfnamefont#1{#1}\fi
\expandafter\ifx\csname citenamefont\endcsname\relax
  \def\citenamefont#1{#1}\fi
\expandafter\ifx\csname url\endcsname\relax
  \def\url#1{\texttt{#1}}\fi
\expandafter\ifx\csname urlprefix\endcsname\relax\def\urlprefix{URL }\fi
\providecommand{\bibinfo}[2]{#2}
\providecommand{\eprint}[2][]{\url{#2}}

\bibitem[{\citenamefont{Fiebig}(2005)}]{Fiebig2005}
\bibinfo{author}{\bibfnamefont{M.}~\bibnamefont{Fiebig}}, \bibinfo{journal}{J.
  Phys. D} \textbf{\bibinfo{volume}{38}}, \bibinfo{pages}{R1}
  (\bibinfo{year}{2005}).

\bibitem[{\citenamefont{Spaldin and Fiebig}(2005)}]{Spaldin2005}
\bibinfo{author}{\bibfnamefont{N.}~\bibnamefont{Spaldin}} \bibnamefont{and}
  \bibinfo{author}{\bibfnamefont{M.}~\bibnamefont{Fiebig}},
  \bibinfo{journal}{Science} \textbf{\bibinfo{volume}{309}},
  \bibinfo{pages}{391} (\bibinfo{year}{2005}).

\bibitem[{\citenamefont{Eerenstein et~al.}(2006)\citenamefont{Eerenstein,
  Mathur, and Scott}}]{Mathur2006}
\bibinfo{author}{\bibfnamefont{W.}~\bibnamefont{Eerenstein}},
  \bibinfo{author}{\bibfnamefont{N.~D.} \bibnamefont{Mathur}},
  \bibnamefont{and} \bibinfo{author}{\bibfnamefont{J.~F.} \bibnamefont{Scott}},
  \bibinfo{journal}{Nature} \textbf{\bibinfo{volume}{442}},
  \bibinfo{pages}{759} (\bibinfo{year}{2006}).

\bibitem[{\citenamefont{Wadley et~al.}(2016)\citenamefont{Wadley, Howells, {\v
  Z}elezn{\'y}, Andrews, Hills, Campion, Nov{\'a}k, Olejn{\'\i}k, Maccherozzi,
  Dhesi et~al.}}]{Wadley2016}
\bibinfo{author}{\bibfnamefont{P.}~\bibnamefont{Wadley}},
  \bibinfo{author}{\bibfnamefont{B.}~\bibnamefont{Howells}},
  \bibinfo{author}{\bibfnamefont{J.}~\bibnamefont{{\v Z}elezn{\'y}}},
  \bibinfo{author}{\bibfnamefont{C.}~\bibnamefont{Andrews}},
  \bibinfo{author}{\bibfnamefont{V.}~\bibnamefont{Hills}},
  \bibinfo{author}{\bibfnamefont{R.~P.} \bibnamefont{Campion}},
  \bibinfo{author}{\bibfnamefont{V.}~\bibnamefont{Nov{\'a}k}},
  \bibinfo{author}{\bibfnamefont{K.}~\bibnamefont{Olejn{\'\i}k}},
  \bibinfo{author}{\bibfnamefont{F.}~\bibnamefont{Maccherozzi}},
  \bibinfo{author}{\bibfnamefont{S.~S.} \bibnamefont{Dhesi}},
  \bibnamefont{et~al.}, \bibinfo{journal}{Science}
  \textbf{\bibinfo{volume}{351}}, \bibinfo{pages}{587} (\bibinfo{year}{2016}).

\bibitem[{\citenamefont{Zener}(1951)}]{Zener1951}
\bibinfo{author}{\bibfnamefont{C.}~\bibnamefont{Zener}},
  \bibinfo{journal}{Phys. Rev.} \textbf{\bibinfo{volume}{82}},
  \bibinfo{pages}{403} (\bibinfo{year}{1951}).

\bibitem[{\citenamefont{de~Gennes}(1960)}]{deGennes1960}
\bibinfo{author}{\bibfnamefont{P.~G.} \bibnamefont{de~Gennes}},
  \bibinfo{journal}{Phys. Rev.} \textbf{\bibinfo{volume}{118}},
  \bibinfo{pages}{141} (\bibinfo{year}{1960}).

\bibitem[{\citenamefont{Kramers}(1934)}]{KRAMERS1934}
\bibinfo{author}{\bibfnamefont{H.}~\bibnamefont{Kramers}},
  \bibinfo{journal}{Physica} \textbf{\bibinfo{volume}{1}}, \bibinfo{pages}{182
  } (\bibinfo{year}{1934}).

\bibitem[{\citenamefont{Goodenough}(1955)}]{Goodenough1955}
\bibinfo{author}{\bibfnamefont{J.~B.} \bibnamefont{Goodenough}},
  \bibinfo{journal}{Phys. Rev.} \textbf{\bibinfo{volume}{100}},
  \bibinfo{pages}{564} (\bibinfo{year}{1955}).

\bibitem[{\citenamefont{Meevasana et~al.}(2011)\citenamefont{Meevasana, King,
  He, Mo, Hashimoto, Tamai, Songsiriritthigul, Baumberger, and
  Sh}}]{Meevasana2011}
\bibinfo{author}{\bibfnamefont{W.}~\bibnamefont{Meevasana}},
  \bibinfo{author}{\bibfnamefont{P.~D.~C.} \bibnamefont{King}},
  \bibinfo{author}{\bibfnamefont{R.~H.} \bibnamefont{He}},
  \bibinfo{author}{\bibfnamefont{S.-K.} \bibnamefont{Mo}},
  \bibinfo{author}{\bibfnamefont{M.}~\bibnamefont{Hashimoto}},
  \bibinfo{author}{\bibfnamefont{A.}~\bibnamefont{Tamai}},
  \bibinfo{author}{\bibfnamefont{P.}~\bibnamefont{Songsiriritthigul}},
  \bibinfo{author}{\bibfnamefont{F.}~\bibnamefont{Baumberger}},
  \bibnamefont{and} \bibinfo{author}{\bibfnamefont{Z.-X.} \bibnamefont{Sh}},
  \bibinfo{journal}{Nat. Mat.} \textbf{\bibinfo{volume}{10}},
  \bibinfo{pages}{114} (\bibinfo{year}{2011}).

\bibitem[{\citenamefont{Ohtomo and Hwang}(2004)}]{Ohtomo2004}
\bibinfo{author}{\bibfnamefont{A.}~\bibnamefont{Ohtomo}} \bibnamefont{and}
  \bibinfo{author}{\bibfnamefont{H.}~\bibnamefont{Hwang}},
  \bibinfo{journal}{Nature} \textbf{\bibinfo{volume}{427}},
  \bibinfo{pages}{423} (\bibinfo{year}{2004}).

\bibitem[{\citenamefont{Reyren et~al.}(2007)\citenamefont{Reyren, Thiel,
  Caviglia, Kourkoutis, Hammerl, Richter, Schneider, Kopp, R{\"u}etschi,
  Jaccard et~al.}}]{2007-Reyren}
\bibinfo{author}{\bibfnamefont{N.}~\bibnamefont{Reyren}},
  \bibinfo{author}{\bibfnamefont{S.}~\bibnamefont{Thiel}},
  \bibinfo{author}{\bibfnamefont{A.~D.} \bibnamefont{Caviglia}},
  \bibinfo{author}{\bibfnamefont{L.~F.} \bibnamefont{Kourkoutis}},
  \bibinfo{author}{\bibfnamefont{G.}~\bibnamefont{Hammerl}},
  \bibinfo{author}{\bibfnamefont{C.}~\bibnamefont{Richter}},
  \bibinfo{author}{\bibfnamefont{C.~W.} \bibnamefont{Schneider}},
  \bibinfo{author}{\bibfnamefont{T.}~\bibnamefont{Kopp}},
  \bibinfo{author}{\bibfnamefont{A.-S.} \bibnamefont{R{\"u}etschi}},
  \bibinfo{author}{\bibfnamefont{D.}~\bibnamefont{Jaccard}},
  \bibnamefont{et~al.}, \bibinfo{journal}{Science}
  \textbf{\bibinfo{volume}{317}}, \bibinfo{pages}{1196} (\bibinfo{year}{2007}).

\bibitem[{\citenamefont{Brinkman et~al.}(2007)\citenamefont{Brinkman, Huijben,
  van Zalk, Huijben, Zeitler, Maan, van~der Wiel, Rijnders, Blank, and
  Hilgenkamp}}]{Brinkman2007}
\bibinfo{author}{\bibfnamefont{A.}~\bibnamefont{Brinkman}},
  \bibinfo{author}{\bibfnamefont{M.}~\bibnamefont{Huijben}},
  \bibinfo{author}{\bibfnamefont{M.}~\bibnamefont{van Zalk}},
  \bibinfo{author}{\bibfnamefont{J.}~\bibnamefont{Huijben}},
  \bibinfo{author}{\bibfnamefont{U.}~\bibnamefont{Zeitler}},
  \bibinfo{author}{\bibfnamefont{J.~C.} \bibnamefont{Maan}},
  \bibinfo{author}{\bibfnamefont{W.~G.} \bibnamefont{van~der Wiel}},
  \bibinfo{author}{\bibfnamefont{G.}~\bibnamefont{Rijnders}},
  \bibinfo{author}{\bibfnamefont{D.~H.~A.} \bibnamefont{Blank}},
  \bibnamefont{and}
  \bibinfo{author}{\bibfnamefont{H.}~\bibnamefont{Hilgenkamp}},
  \bibinfo{journal}{Nature Materials} \textbf{\bibinfo{volume}{6}},
  \bibinfo{pages}{493–496} (\bibinfo{year}{2007}).

\bibitem[{\citenamefont{Stengel}(2011)}]{Stengel2011}
\bibinfo{author}{\bibfnamefont{M.}~\bibnamefont{Stengel}},
  \bibinfo{journal}{Phys. Rev. Lett.} \textbf{\bibinfo{volume}{106}},
  \bibinfo{pages}{136803} (\bibinfo{year}{2011}).

\bibitem[{\citenamefont{Niu et~al.}(2017)\citenamefont{Niu, Zhang, Gan,
  Christensen, Soosten, Garcia-Suarez, Riisager, Wang, Xu, Zhang
  et~al.}}]{Niu2017}
\bibinfo{author}{\bibfnamefont{W.}~\bibnamefont{Niu}},
  \bibinfo{author}{\bibfnamefont{Y.}~\bibnamefont{Zhang}},
  \bibinfo{author}{\bibfnamefont{Y.}~\bibnamefont{Gan}},
  \bibinfo{author}{\bibfnamefont{D.~V.} \bibnamefont{Christensen}},
  \bibinfo{author}{\bibfnamefont{M.~V.} \bibnamefont{Soosten}},
  \bibinfo{author}{\bibfnamefont{E.~J.} \bibnamefont{Garcia-Suarez}},
  \bibinfo{author}{\bibfnamefont{A.}~\bibnamefont{Riisager}},
  \bibinfo{author}{\bibfnamefont{X.}~\bibnamefont{Wang}},
  \bibinfo{author}{\bibfnamefont{Y.}~\bibnamefont{Xu}},
  \bibinfo{author}{\bibfnamefont{R.}~\bibnamefont{Zhang}},
  \bibnamefont{et~al.}, \bibinfo{journal}{Nano Letters}
  \textbf{\bibinfo{volume}{17}}, \bibinfo{pages}{6878} (\bibinfo{year}{2017}).

\bibitem[{\citenamefont{Maznichenko et~al.}(2020)\citenamefont{Maznichenko,
  Ostanin, Ernst, Henk, and Mertig}}]{Mertig2020}
\bibinfo{author}{\bibfnamefont{I.~V.} \bibnamefont{Maznichenko}},
  \bibinfo{author}{\bibfnamefont{S.}~\bibnamefont{Ostanin}},
  \bibinfo{author}{\bibfnamefont{A.}~\bibnamefont{Ernst}},
  \bibinfo{author}{\bibfnamefont{J.}~\bibnamefont{Henk}}, \bibnamefont{and}
  \bibinfo{author}{\bibfnamefont{I.}~\bibnamefont{Mertig}},
  \bibinfo{journal}{Phys. Status Solidi B} \textbf{\bibinfo{volume}{257}},
  \bibinfo{pages}{1900540} (\bibinfo{year}{2020}).

\bibitem[{\citenamefont{Vildosola et~al.}(2013)\citenamefont{Vildosola,
  Güller, and Llois}}]{Vildosola2013}
\bibinfo{author}{\bibfnamefont{V.}~\bibnamefont{Vildosola}},
  \bibinfo{author}{\bibfnamefont{F.}~\bibnamefont{Güller}}, \bibnamefont{and}
  \bibinfo{author}{\bibfnamefont{A.~M.} \bibnamefont{Llois}},
  \bibinfo{journal}{Physical Review Letters} \textbf{\bibinfo{volume}{110}},
  \bibinfo{pages}{206805} (\bibinfo{year}{2013}).

\bibitem[{\citenamefont{Meir et~al.}(2017)\citenamefont{Meir, Gorol, Kopp, and
  Hammerl}}]{Meir2017}
\bibinfo{author}{\bibfnamefont{B.}~\bibnamefont{Meir}},
  \bibinfo{author}{\bibfnamefont{S.}~\bibnamefont{Gorol}},
  \bibinfo{author}{\bibfnamefont{T.}~\bibnamefont{Kopp}}, \bibnamefont{and}
  \bibinfo{author}{\bibfnamefont{G.}~\bibnamefont{Hammerl}},
  \bibinfo{journal}{Phys. Rev. B} \textbf{\bibinfo{volume}{96}},
  \bibinfo{pages}{R100507} (\bibinfo{year}{2017}).

\bibitem[{\citenamefont{Di~Napoli et~al.}(2021)\citenamefont{Di~Napoli, Helman,
  Llois, and Vildosola}}]{Dinapoli2021}
\bibinfo{author}{\bibfnamefont{S.}~\bibnamefont{Di~Napoli}},
  \bibinfo{author}{\bibfnamefont{C.}~\bibnamefont{Helman}},
  \bibinfo{author}{\bibfnamefont{A.~M.} \bibnamefont{Llois}}, \bibnamefont{and}
  \bibinfo{author}{\bibfnamefont{V.}~\bibnamefont{Vildosola}},
  \bibinfo{journal}{Phys. Rev. B} \textbf{\bibinfo{volume}{103}},
  \bibinfo{pages}{174509} (\bibinfo{year}{2021}).

\bibitem[{\citenamefont{Borisov et~al.}(2015)\citenamefont{Borisov, Ostanin,
  and Mertig}}]{2015-Mertig}
\bibinfo{author}{\bibfnamefont{V.}~\bibnamefont{Borisov}},
  \bibinfo{author}{\bibfnamefont{S.}~\bibnamefont{Ostanin}}, \bibnamefont{and}
  \bibinfo{author}{\bibfnamefont{I.}~\bibnamefont{Mertig}},
  \bibinfo{journal}{Phys. Chem. Chem. Phys.} \textbf{\bibinfo{volume}{17}},
  \bibinfo{pages}{12812} (\bibinfo{year}{2015}).

\bibitem[{\citenamefont{Weng et~al.}(2021)\citenamefont{Weng, Niu, Huang, An,
  and Dong}}]{Weng2021}
\bibinfo{author}{\bibfnamefont{Y.}~\bibnamefont{Weng}},
  \bibinfo{author}{\bibfnamefont{W.}~\bibnamefont{Niu}},
  \bibinfo{author}{\bibfnamefont{X.}~\bibnamefont{Huang}},
  \bibinfo{author}{\bibfnamefont{M.}~\bibnamefont{An}}, \bibnamefont{and}
  \bibinfo{author}{\bibfnamefont{S.}~\bibnamefont{Dong}},
  \bibinfo{journal}{Phys. Rev. B} \textbf{\bibinfo{volume}{103}},
  \bibinfo{pages}{214101} (\bibinfo{year}{2021}).

\bibitem[{\citenamefont{Aggoune and Draxl}(2021)}]{Draxl2021}
\bibinfo{author}{\bibfnamefont{W.}~\bibnamefont{Aggoune}} \bibnamefont{and}
  \bibinfo{author}{\bibfnamefont{C.}~\bibnamefont{Draxl}},
  \bibinfo{journal}{npj Computational Materials} \textbf{\bibinfo{volume}{7}},
  \bibinfo{pages}{174} (\bibinfo{year}{2021}).

\bibitem[{\citenamefont{Fang et~al.}(2023)\citenamefont{Fang, Aggoune, Ren, and
  Draxl}}]{2023Draxl}
\bibinfo{author}{\bibfnamefont{L.}~\bibnamefont{Fang}},
  \bibinfo{author}{\bibfnamefont{W.}~\bibnamefont{Aggoune}},
  \bibinfo{author}{\bibfnamefont{W.}~\bibnamefont{Ren}}, \bibnamefont{and}
  \bibinfo{author}{\bibfnamefont{C.}~\bibnamefont{Draxl}},
  \bibinfo{journal}{ACS Applied Materials \& Interfaces}
  \textbf{\bibinfo{volume}{15}}, \bibinfo{pages}{11314} (\bibinfo{year}{2023}),
  \bibinfo{note}{pMID: 36787465}.

\bibitem[{\citenamefont{Yin et~al.}(2015)\citenamefont{Yin, Aguado-Puente, Qu,
  and Artacho}}]{Artacho2015}
\bibinfo{author}{\bibfnamefont{B.}~\bibnamefont{Yin}},
  \bibinfo{author}{\bibfnamefont{P.}~\bibnamefont{Aguado-Puente}},
  \bibinfo{author}{\bibfnamefont{S.}~\bibnamefont{Qu}}, \bibnamefont{and}
  \bibinfo{author}{\bibfnamefont{E.}~\bibnamefont{Artacho}},
  \bibinfo{journal}{Phys. Rev. B} \textbf{\bibinfo{volume}{92}},
  \bibinfo{pages}{115406} (\bibinfo{year}{2015}).

\bibitem[{\citenamefont{Pedroso et~al.}(2020)\citenamefont{Pedroso, Barral,
  Graf, Llois, Aguirre, Steren, and Di~Napoli}}]{2020-Pedroso}
\bibinfo{author}{\bibfnamefont{A.~L.} \bibnamefont{Pedroso}},
  \bibinfo{author}{\bibfnamefont{M.~A.} \bibnamefont{Barral}},
  \bibinfo{author}{\bibfnamefont{M.~E.} \bibnamefont{Graf}},
  \bibinfo{author}{\bibfnamefont{A.~M.} \bibnamefont{Llois}},
  \bibinfo{author}{\bibfnamefont{M.~H.} \bibnamefont{Aguirre}},
  \bibinfo{author}{\bibfnamefont{L.~B.} \bibnamefont{Steren}},
  \bibnamefont{and}
  \bibinfo{author}{\bibfnamefont{S.}~\bibnamefont{Di~Napoli}},
  \bibinfo{journal}{Phys. Rev. B} \textbf{\bibinfo{volume}{102}},
  \bibinfo{pages}{085432} (\bibinfo{year}{2020}).

\bibitem[{\citenamefont{Duan et~al.}(2006)\citenamefont{Duan, Jaswal, and
  Tsymbal}}]{Tsymbal2006}
\bibinfo{author}{\bibfnamefont{C.-G.} \bibnamefont{Duan}},
  \bibinfo{author}{\bibfnamefont{S.~S.} \bibnamefont{Jaswal}},
  \bibnamefont{and} \bibinfo{author}{\bibfnamefont{E.~Y.}
  \bibnamefont{Tsymbal}}, \bibinfo{journal}{Phys. Rev. Lett.}
  \textbf{\bibinfo{volume}{97}}, \bibinfo{pages}{047201}
  (\bibinfo{year}{2006}).

\bibitem[{\citenamefont{Niranjan et~al.}(2008)\citenamefont{Niranjan, Velev,
  Duan, Jaswal, and Tsymbal}}]{Tsymbal2008}
\bibinfo{author}{\bibfnamefont{M.~K.} \bibnamefont{Niranjan}},
  \bibinfo{author}{\bibfnamefont{J.~P.} \bibnamefont{Velev}},
  \bibinfo{author}{\bibfnamefont{C.-G.} \bibnamefont{Duan}},
  \bibinfo{author}{\bibfnamefont{S.~S.} \bibnamefont{Jaswal}},
  \bibnamefont{and} \bibinfo{author}{\bibfnamefont{E.~Y.}
  \bibnamefont{Tsymbal}}, \bibinfo{journal}{Phys. Rev. B}
  \textbf{\bibinfo{volume}{78}}, \bibinfo{pages}{104405}
  (\bibinfo{year}{2008}).

\bibitem[{\citenamefont{Niranjan et~al.}(2010)\citenamefont{Niranjan, Duan,
  Jaswal, and Tsymbal}}]{Tsymbal2010}
\bibinfo{author}{\bibfnamefont{M.~K.} \bibnamefont{Niranjan}},
  \bibinfo{author}{\bibfnamefont{C.-G.} \bibnamefont{Duan}},
  \bibinfo{author}{\bibfnamefont{S.~S.} \bibnamefont{Jaswal}},
  \bibnamefont{and} \bibinfo{author}{\bibfnamefont{E.~Y.}
  \bibnamefont{Tsymbal}}, \bibinfo{journal}{Applied Physics Letters}
  \textbf{\bibinfo{volume}{96}}, \bibinfo{pages}{222504}
  (\bibinfo{year}{2010}).

\bibitem[{\citenamefont{Burton and Tsymbal}(2009)}]{Burton2009}
\bibinfo{author}{\bibfnamefont{J.~D.} \bibnamefont{Burton}} \bibnamefont{and}
  \bibinfo{author}{\bibfnamefont{E.~Y.} \bibnamefont{Tsymbal}},
  \bibinfo{journal}{Phys. Rev. B} \textbf{\bibinfo{volume}{80}},
  \bibinfo{pages}{174406} (\bibinfo{year}{2009}).

\bibitem[{\citenamefont{Borek et~al.}(2012)\citenamefont{Borek, Maznichenko,
  Fischer, Hergert, Mertig, Ernst, Ostanin, and Chass\'e}}]{Mertig2012}
\bibinfo{author}{\bibfnamefont{S.}~\bibnamefont{Borek}},
  \bibinfo{author}{\bibfnamefont{I.~V.} \bibnamefont{Maznichenko}},
  \bibinfo{author}{\bibfnamefont{G.}~\bibnamefont{Fischer}},
  \bibinfo{author}{\bibfnamefont{W.}~\bibnamefont{Hergert}},
  \bibinfo{author}{\bibfnamefont{I.}~\bibnamefont{Mertig}},
  \bibinfo{author}{\bibfnamefont{A.}~\bibnamefont{Ernst}},
  \bibinfo{author}{\bibfnamefont{S.}~\bibnamefont{Ostanin}}, \bibnamefont{and}
  \bibinfo{author}{\bibfnamefont{A.}~\bibnamefont{Chass\'e}},
  \bibinfo{journal}{Phys. Rev. B} \textbf{\bibinfo{volume}{85}},
  \bibinfo{pages}{134432} (\bibinfo{year}{2012}).

\bibitem[{\citenamefont{Wieder}(1955)}]{1955-P_BTO}
\bibinfo{author}{\bibfnamefont{H.~H.} \bibnamefont{Wieder}},
  \bibinfo{journal}{Phys. Rev.} \textbf{\bibinfo{volume}{99}},
  \bibinfo{pages}{1161} (\bibinfo{year}{1955}).

\bibitem[{\citenamefont{Bl\"ochl}(1994)}]{PAW}
\bibinfo{author}{\bibfnamefont{P.}~\bibnamefont{Bl\"ochl}},
  \bibinfo{journal}{Phys. Rev. B} \textbf{\bibinfo{volume}{50}},
  \bibinfo{pages}{17953} (\bibinfo{year}{1994}).

\bibitem[{\citenamefont{Kresse and Furthm\"uller}(1996)}]{VASP}
\bibinfo{author}{\bibfnamefont{G.}~\bibnamefont{Kresse}} \bibnamefont{and}
  \bibinfo{author}{\bibfnamefont{J.}~\bibnamefont{Furthm\"uller}},
  \bibinfo{journal}{Phys. Rev. B} \textbf{\bibinfo{volume}{54}},
  \bibinfo{pages}{11169} (\bibinfo{year}{1996}).

\bibitem[{\citenamefont{Kresse and Joubert}(1999)}]{PAW-VASP}
\bibinfo{author}{\bibfnamefont{G.}~\bibnamefont{Kresse}} \bibnamefont{and}
  \bibinfo{author}{\bibfnamefont{D.}~\bibnamefont{Joubert}},
  \bibinfo{journal}{Phys. Rev. B} \textbf{\bibinfo{volume}{59}},
  \bibinfo{pages}{1758} (\bibinfo{year}{1999}).

\bibitem[{\citenamefont{Perdew and Zunger}(1981)}]{LDA1}
\bibinfo{author}{\bibfnamefont{J.~P.} \bibnamefont{Perdew}} \bibnamefont{and}
  \bibinfo{author}{\bibfnamefont{A.}~\bibnamefont{Zunger}},
  \bibinfo{journal}{Phys. Rev. B} \textbf{\bibinfo{volume}{23}},
  \bibinfo{pages}{5048} (\bibinfo{year}{1981}).

\bibitem[{\citenamefont{Ceperley and Alder}(1980)}]{LDA2}
\bibinfo{author}{\bibfnamefont{D.}~\bibnamefont{Ceperley}} \bibnamefont{and}
  \bibinfo{author}{\bibfnamefont{B.}~\bibnamefont{Alder}},
  \bibinfo{journal}{Phys. Rev. Lett.} \textbf{\bibinfo{volume}{45}},
  \bibinfo{pages}{466} (\bibinfo{year}{1980}).

\bibitem[{\citenamefont{Filippetti and Pickett}(1999)}]{Picket1999}
\bibinfo{author}{\bibfnamefont{A.}~\bibnamefont{Filippetti}} \bibnamefont{and}
  \bibinfo{author}{\bibfnamefont{W.}~\bibnamefont{Pickett}},
  \bibinfo{journal}{Phys. Rev. Lett.} \textbf{\bibinfo{volume}{83}},
  \bibinfo{pages}{4184} (\bibinfo{year}{1999}).

\bibitem[{\citenamefont{Filippetti and Pickett}(2000)}]{Picket2000}
\bibinfo{author}{\bibfnamefont{A.}~\bibnamefont{Filippetti}} \bibnamefont{and}
  \bibinfo{author}{\bibfnamefont{W.}~\bibnamefont{Pickett}},
  \bibinfo{journal}{Phys. Rev. B} \textbf{\bibinfo{volume}{62}},
  \bibinfo{pages}{11571} (\bibinfo{year}{2000}).

\bibitem[{\citenamefont{Keshavarz et~al.}(2017)\citenamefont{Keshavarz,
  Kvashnin, Rodrigues, Pereiro, Di~Marco, Autieri, Nordstr\"om, Solovyev,
  Sanyal, and Eriksson}}]{Nordstrom2017}
\bibinfo{author}{\bibfnamefont{S.}~\bibnamefont{Keshavarz}},
  \bibinfo{author}{\bibfnamefont{Y.~O.} \bibnamefont{Kvashnin}},
  \bibinfo{author}{\bibfnamefont{D.~C.~M.} \bibnamefont{Rodrigues}},
  \bibinfo{author}{\bibfnamefont{M.}~\bibnamefont{Pereiro}},
  \bibinfo{author}{\bibfnamefont{I.}~\bibnamefont{Di~Marco}},
  \bibinfo{author}{\bibfnamefont{C.}~\bibnamefont{Autieri}},
  \bibinfo{author}{\bibfnamefont{L.}~\bibnamefont{Nordstr\"om}},
  \bibinfo{author}{\bibfnamefont{I.~V.} \bibnamefont{Solovyev}},
  \bibinfo{author}{\bibfnamefont{B.}~\bibnamefont{Sanyal}}, \bibnamefont{and}
  \bibinfo{author}{\bibfnamefont{O.}~\bibnamefont{Eriksson}},
  \bibinfo{journal}{Phys. Rev. B} \textbf{\bibinfo{volume}{95}},
  \bibinfo{pages}{115120} (\bibinfo{year}{2017}).

\bibitem[{\citenamefont{Neaton et~al.}(2002)\citenamefont{Neaton, Hsueh, and
  Rabe}}]{Rabe2002}
\bibinfo{author}{\bibfnamefont{J.}~\bibnamefont{Neaton}},
  \bibinfo{author}{\bibfnamefont{C.-L.} \bibnamefont{Hsueh}}, \bibnamefont{and}
  \bibinfo{author}{\bibfnamefont{K.}~\bibnamefont{Rabe}},
  \bibinfo{journal}{Mat. Res. Soc. Symp. Proc.} \textbf{\bibinfo{volume}{718}},
  \bibinfo{pages}{D10.26} (\bibinfo{year}{2002}).

\bibitem[{\citenamefont{Neaton and Rabe}(2003)}]{Rabe2003}
\bibinfo{author}{\bibfnamefont{J.~B.} \bibnamefont{Neaton}} \bibnamefont{and}
  \bibinfo{author}{\bibfnamefont{K.~M.} \bibnamefont{Rabe}},
  \bibinfo{journal}{Applied Physics Letters} \textbf{\bibinfo{volume}{82}},
  \bibinfo{pages}{1586} (\bibinfo{year}{2003}).

\bibitem[{\citenamefont{Zhang et~al.}(2017)\citenamefont{Zhang, Sun, Perdew,
  and Wu}}]{PhysRevB.96.035143}
\bibinfo{author}{\bibfnamefont{Y.}~\bibnamefont{Zhang}},
  \bibinfo{author}{\bibfnamefont{J.}~\bibnamefont{Sun}},
  \bibinfo{author}{\bibfnamefont{J.~P.} \bibnamefont{Perdew}},
  \bibnamefont{and} \bibinfo{author}{\bibfnamefont{X.}~\bibnamefont{Wu}},
  \bibinfo{journal}{Phys. Rev. B} \textbf{\bibinfo{volume}{96}},
  \bibinfo{pages}{035143} (\bibinfo{year}{2017}).

\bibitem[{\citenamefont{Liechtenstein et~al.}(1995)\citenamefont{Liechtenstein,
  Anisimov, and Zaanen}}]{Liechtenstein95}
\bibinfo{author}{\bibfnamefont{A.~I.} \bibnamefont{Liechtenstein}},
  \bibinfo{author}{\bibfnamefont{V.~I.} \bibnamefont{Anisimov}},
  \bibnamefont{and} \bibinfo{author}{\bibfnamefont{J.}~\bibnamefont{Zaanen}},
  \bibinfo{journal}{Phys. Rev. B} \textbf{\bibinfo{volume}{52}},
  \bibinfo{pages}{R5467} (\bibinfo{year}{1995}).

\bibitem[{\citenamefont{Loshkareva et~al.}(2004)\citenamefont{Loshkareva,
  Nomerovannaya, Mostovshchikova, Makhnev, Sukhorukov, Solin, Arbuzova, Naumov,
  Kostromitina, Balbashov et~al.}}]{Loshkareva2004}
\bibinfo{author}{\bibfnamefont{N.~N.} \bibnamefont{Loshkareva}},
  \bibinfo{author}{\bibfnamefont{L.~V.} \bibnamefont{Nomerovannaya}},
  \bibinfo{author}{\bibfnamefont{E.~V.} \bibnamefont{Mostovshchikova}},
  \bibinfo{author}{\bibfnamefont{A.~A.} \bibnamefont{Makhnev}},
  \bibinfo{author}{\bibfnamefont{Y.~P.} \bibnamefont{Sukhorukov}},
  \bibinfo{author}{\bibfnamefont{N.~I.} \bibnamefont{Solin}},
  \bibinfo{author}{\bibfnamefont{T.~I.} \bibnamefont{Arbuzova}},
  \bibinfo{author}{\bibfnamefont{S.~V.} \bibnamefont{Naumov}},
  \bibinfo{author}{\bibfnamefont{N.~V.} \bibnamefont{Kostromitina}},
  \bibinfo{author}{\bibfnamefont{A.~M.} \bibnamefont{Balbashov}},
  \bibnamefont{et~al.}, \bibinfo{journal}{Phys. Rev. B}
  \textbf{\bibinfo{volume}{70}}, \bibinfo{pages}{224406}
  (\bibinfo{year}{2004}).

\bibitem[{\citenamefont{Tenne et~al.}(2006)\citenamefont{Tenne, Bruchhausen,
  Lanzillotti-Kimura, Fainstein, Katiyar, Cantarero, Soukiassian,
  Vaithyanathan, Haeni, Tian et~al.}}]{Rabe2006}
\bibinfo{author}{\bibfnamefont{D.~A.} \bibnamefont{Tenne}},
  \bibinfo{author}{\bibfnamefont{A.}~\bibnamefont{Bruchhausen}},
  \bibinfo{author}{\bibfnamefont{N.~D.} \bibnamefont{Lanzillotti-Kimura}},
  \bibinfo{author}{\bibfnamefont{A.}~\bibnamefont{Fainstein}},
  \bibinfo{author}{\bibfnamefont{R.~S.} \bibnamefont{Katiyar}},
  \bibinfo{author}{\bibfnamefont{A.}~\bibnamefont{Cantarero}},
  \bibinfo{author}{\bibfnamefont{A.}~\bibnamefont{Soukiassian}},
  \bibinfo{author}{\bibfnamefont{V.}~\bibnamefont{Vaithyanathan}},
  \bibinfo{author}{\bibfnamefont{J.~H.} \bibnamefont{Haeni}},
  \bibinfo{author}{\bibfnamefont{W.}~\bibnamefont{Tian}}, \bibnamefont{et~al.},
  \bibinfo{journal}{Science} \textbf{\bibinfo{volume}{313}},
  \bibinfo{pages}{1614} (\bibinfo{year}{2006}).

\bibitem[{\citenamefont{Gerra et~al.}(2006)\citenamefont{Gerra, Tagantsev,
  Setter, and Parlinski}}]{Gerra2006}
\bibinfo{author}{\bibfnamefont{G.}~\bibnamefont{Gerra}},
  \bibinfo{author}{\bibfnamefont{A.~K.} \bibnamefont{Tagantsev}},
  \bibinfo{author}{\bibfnamefont{N.}~\bibnamefont{Setter}}, \bibnamefont{and}
  \bibinfo{author}{\bibfnamefont{K.}~\bibnamefont{Parlinski}},
  \bibinfo{journal}{Phys. Rev. Lett.} \textbf{\bibinfo{volume}{96}},
  \bibinfo{pages}{107603} (\bibinfo{year}{2006}).

\bibitem[{\citenamefont{Fechner et~al.}(2008)\citenamefont{Fechner, Ostanin,
  and Mertig}}]{Mertig2008}
\bibinfo{author}{\bibfnamefont{M.}~\bibnamefont{Fechner}},
  \bibinfo{author}{\bibfnamefont{S.}~\bibnamefont{Ostanin}}, \bibnamefont{and}
  \bibinfo{author}{\bibfnamefont{I.}~\bibnamefont{Mertig}},
  \bibinfo{journal}{Phys. Rev. B} \textbf{\bibinfo{volume}{77}},
  \bibinfo{pages}{094112} (\bibinfo{year}{2008}).

\bibitem[{\citenamefont{King-Smith and Vanderbilt}(1993)}]{Vanderbilt1993}
\bibinfo{author}{\bibfnamefont{R.~D.} \bibnamefont{King-Smith}}
  \bibnamefont{and}
  \bibinfo{author}{\bibfnamefont{D.}~\bibnamefont{Vanderbilt}},
  \bibinfo{journal}{Phys. Rev. B} \textbf{\bibinfo{volume}{47}},
  \bibinfo{pages}{1651} (\bibinfo{year}{1993}).

\bibitem[{\citenamefont{Momma and Izumi}(2008)}]{VESTA}
\bibinfo{author}{\bibfnamefont{K.}~\bibnamefont{Momma}} \bibnamefont{and}
  \bibinfo{author}{\bibfnamefont{F.}~\bibnamefont{Izumi}},
  \bibinfo{journal}{Journal of Applied Crystallography}
  \textbf{\bibinfo{volume}{41}}, \bibinfo{pages}{653} (\bibinfo{year}{2008}).

\bibitem[{\citenamefont{Bousquet and Spaldin}(2011)}]{Spaldin2011}
\bibinfo{author}{\bibfnamefont{E.}~\bibnamefont{Bousquet}} \bibnamefont{and}
  \bibinfo{author}{\bibfnamefont{N.}~\bibnamefont{Spaldin}},
  \bibinfo{journal}{Phys. Rev. Lett.} \textbf{\bibinfo{volume}{107}},
  \bibinfo{pages}{197603} (\bibinfo{year}{2011}).

\bibitem[{\citenamefont{Vistoli et~al.}(2019)\citenamefont{Vistoli, Wang,
  Sander, Zhu, Casals, Cichelero, Barthélémy, Fusil, Herranz, Valencia
  et~al.}}]{Bibes2019}
\bibinfo{author}{\bibfnamefont{L.}~\bibnamefont{Vistoli}},
  \bibinfo{author}{\bibfnamefont{W.}~\bibnamefont{Wang}},
  \bibinfo{author}{\bibfnamefont{A.}~\bibnamefont{Sander}},
  \bibinfo{author}{\bibfnamefont{Q.}~\bibnamefont{Zhu}},
  \bibinfo{author}{\bibfnamefont{B.}~\bibnamefont{Casals}},
  \bibinfo{author}{\bibfnamefont{R.}~\bibnamefont{Cichelero}},
  \bibinfo{author}{\bibfnamefont{A.}~\bibnamefont{Barthélémy}},
  \bibinfo{author}{\bibfnamefont{S.}~\bibnamefont{Fusil}},
  \bibinfo{author}{\bibfnamefont{G.}~\bibnamefont{Herranz}},
  \bibinfo{author}{\bibfnamefont{S.}~\bibnamefont{Valencia}},
  \bibnamefont{et~al.}, \bibinfo{journal}{Nature Phys.}
  \textbf{\bibinfo{volume}{15}}, \bibinfo{pages}{67} (\bibinfo{year}{2019}).

\bibitem[{\citenamefont{Kwei et~al.}(1993)\citenamefont{Kwei, Lawson, Billinge,
  , and Cheong}}]{Kwei1993}
\bibinfo{author}{\bibfnamefont{G.~H.} \bibnamefont{Kwei}},
  \bibinfo{author}{\bibfnamefont{A.~C.} \bibnamefont{Lawson}},
  \bibinfo{author}{\bibfnamefont{S.~J.~L.} \bibnamefont{Billinge}}, ,
  \bibnamefont{and} \bibinfo{author}{\bibfnamefont{S.~W.}
  \bibnamefont{Cheong}}, \bibinfo{journal}{J. Phys. Chem.}
  \textbf{\bibinfo{volume}{97}}, \bibinfo{pages}{2368} (\bibinfo{year}{1993}).

\end{thebibliography}


\begin{thebibliography}{4}
\expandafter\ifx\csname natexlab\endcsname\relax\def\natexlab#1{#1}\fi
\expandafter\ifx\csname bibnamefont\endcsname\relax
  \def\bibnamefont#1{#1}\fi
\expandafter\ifx\csname bibfnamefont\endcsname\relax
  \def\bibfnamefont#1{#1}\fi
\expandafter\ifx\csname citenamefont\endcsname\relax
  \def\citenamefont#1{#1}\fi
\expandafter\ifx\csname url\endcsname\relax
  \def\url#1{\texttt{#1}}\fi
\expandafter\ifx\csname urlprefix\endcsname\relax\def\urlprefix{URL }\fi
\providecommand{\bibinfo}[2]{#2}
\providecommand{\eprint}[2][]{\url{#2}}

\bibitem[{\citenamefont{Kim et~al.}(2021)\citenamefont{Kim, Rossell, Campanini,
  Erni, Puigmart{\'\i}-Luis, Chen, and Pan{\'e}}}]{kim2021}
\bibinfo{author}{\bibfnamefont{D.}~\bibnamefont{Kim}},
  \bibinfo{author}{\bibfnamefont{M.~D.} \bibnamefont{Rossell}},
  \bibinfo{author}{\bibfnamefont{M.}~\bibnamefont{Campanini}},
  \bibinfo{author}{\bibfnamefont{R.}~\bibnamefont{Erni}},
  \bibinfo{author}{\bibfnamefont{J.}~\bibnamefont{Puigmart{\'\i}-Luis}},
  \bibinfo{author}{\bibfnamefont{X.-Z.} \bibnamefont{Chen}}, \bibnamefont{and}
  \bibinfo{author}{\bibfnamefont{S.}~\bibnamefont{Pan{\'e}}},
  \bibinfo{journal}{Applied Physics Letters} \textbf{\bibinfo{volume}{119}},
  \bibinfo{pages}{012901} (\bibinfo{year}{2021}).

\bibitem[{\citenamefont{Choi et~al.}(2004)\citenamefont{Choi, Biegalski, Li,
  Sharan, Schubert, Uecker, Reiche, Chen, Pan, Gopalan
  et~al.}}]{choi2004enhancement}
\bibinfo{author}{\bibfnamefont{K.~J.} \bibnamefont{Choi}},
  \bibinfo{author}{\bibfnamefont{M.}~\bibnamefont{Biegalski}},
  \bibinfo{author}{\bibfnamefont{Y.}~\bibnamefont{Li}},
  \bibinfo{author}{\bibfnamefont{A.}~\bibnamefont{Sharan}},
  \bibinfo{author}{\bibfnamefont{J.}~\bibnamefont{Schubert}},
  \bibinfo{author}{\bibfnamefont{R.}~\bibnamefont{Uecker}},
  \bibinfo{author}{\bibfnamefont{P.}~\bibnamefont{Reiche}},
  \bibinfo{author}{\bibfnamefont{Y.}~\bibnamefont{Chen}},
  \bibinfo{author}{\bibfnamefont{X.}~\bibnamefont{Pan}},
  \bibinfo{author}{\bibfnamefont{V.}~\bibnamefont{Gopalan}},
  \bibnamefont{et~al.}, \bibinfo{journal}{Science}
  \textbf{\bibinfo{volume}{306}}, \bibinfo{pages}{1005} (\bibinfo{year}{2004}).

\bibitem[{\citenamefont{Shebanov}(1981)}]{shebanov1981x}
\bibinfo{author}{\bibfnamefont{L.}~\bibnamefont{Shebanov}},
  \bibinfo{journal}{physica status solidi (a)} \textbf{\bibinfo{volume}{65}},
  \bibinfo{pages}{321} (\bibinfo{year}{1981}).

\bibitem[{\citenamefont{Kholkin et~al.}(2007)\citenamefont{Kholkin, Kalinin,
  Roelofs, and Gruverman}}]{kholkin2007review}
\bibinfo{author}{\bibfnamefont{A.}~\bibnamefont{Kholkin}},
  \bibinfo{author}{\bibfnamefont{S.}~\bibnamefont{Kalinin}},
  \bibinfo{author}{\bibfnamefont{A.}~\bibnamefont{Roelofs}}, \bibnamefont{and}
  \bibinfo{author}{\bibfnamefont{A.}~\bibnamefont{Gruverman}},
  \bibinfo{journal}{Scanning Probe Microscopy: Electrical and Electromechanical
  Phenomena at the Nanoscale} pp. \bibinfo{pages}{173--214}
  (\bibinfo{year}{2007}).

\end{thebibliography}

\end{document}


\title{Supplementary information \\ 
\vspace{1cm}
Magnetic transition and spin-polarized two-dimensional electron gas controlled by polarization switching in strained CaMnO$_3$/BaTiO$_3$ slabs}

\author{S. Di Napoli}
\affiliation{Departamento de F\'{\i}sica de la Materia Condensada, GIyA-CNEA, Avenida
General Paz 1499, (1650) San Mart\'{\i}n, Pcia. de Buenos Aires, Argentina}
\affiliation{Instituto de Nanociencia y Nanotecnolog\'{\i}a (INN CNEA-CONICET), (1650) San Mart\'{\i}n, Pcia. de Buenos Aires, Argentina}

\author{A. Rom\'an}
\affiliation{Laboratorio de Nanoestructuras Magn\'eticas y Dispositivos, Centro At\'omico Constituyentes, 1650 San Mart\'{\i}n, Pcia. de Buenos Aires, Argentina}
\affiliation{Instituto de Nanociencia y Nanotecnolog\'{\i}a (INN CNEA-CONICET), 1650 San Mart\'{\i}n, Argentina}

\author{A.M. Llois}
\affiliation{Departamento de F\'{\i}sica de la Materia Condensada, GIyA-CNEA, Avenida
General Paz 1499, (1650) San Mart\'{\i}n, Pcia. de Buenos Aires, Argentina}
\affiliation{Instituto de Nanociencia y Nanotecnolog\'{\i}a (INN CNEA-CONICET), (1650) San Mart\'{\i}n, Pcia. de Buenos Aires, Argentina}

\author{M.H. Aguirre}
\affiliation{ Instituto de Nanociencia y Materiales de Arag\'on, CSIC, E-50018 Zaragoza, Spain.}
\affiliation{Departamento de F\'{\i}sica de la Materia Condensada, Universidad de Zaragoza, E-50009 Zaragoza, Spain.}
\affiliation{Laboratorio de Microscop\'{\i}as Avanzadas, Universidad de Zaragoza, E-50018 Zaragoza, Spain.}

\author{L.B. Steren}
\affiliation{Laboratorio de Nanoestructuras Magn\'eticas y Dispositivos, Centro At\'omico Constituyentes, 1650 San Mart\'{\i}n, Pcia. de Buenos Aires, Argentina}
\affiliation{Instituto de Nanociencia y Nanotecnolog\'{\i}a (INN CNEA-CONICET), 1650 San Mart\'{\i}n, Argentina}

\author{M.A. Barral}
\affiliation{Departamento de F\'{\i}sica de la Materia Condensada, GIyA-CNEA, Avenida
General Paz 1499, (1650) San Mart\'{\i}n, Pcia. de Buenos Aires, Argentina}
\affiliation{Instituto de Nanociencia y Nanotecnolog\'{\i}a (INN CNEA-CONICET), (1650) San Mart\'{\i}n, Pcia. de Buenos Aires, Argentina}
\email{barral@tandar.cnea.gov.ar}

\maketitle

\section*{Epitaxial growth of BaTiO$_3$ thin film on an SrTiO$_3$ substrate}
\paragraph{Sample fabrication:} A BaTiO$_3$ (BTO) film with a thickness of 60 nm was deposited on an SrTiO$_3$ (STO) substrate using DC magnetron sputtering in a mixed Ar/O$_2$ atmosphere at a chamber pressure of 0.4 torr and a substrate temperature of 720°C.

\paragraph{Structural characterization:} The structural characterization of the bilayer was done with High-Resolution Scanning Transmission Electron Microscopy with High Angular Annular detector (HRSTEM-HAADF) measurements and X-ray diffraction experiments. 
\begin{figure}[h!]
\includegraphics[width=\textwidth]{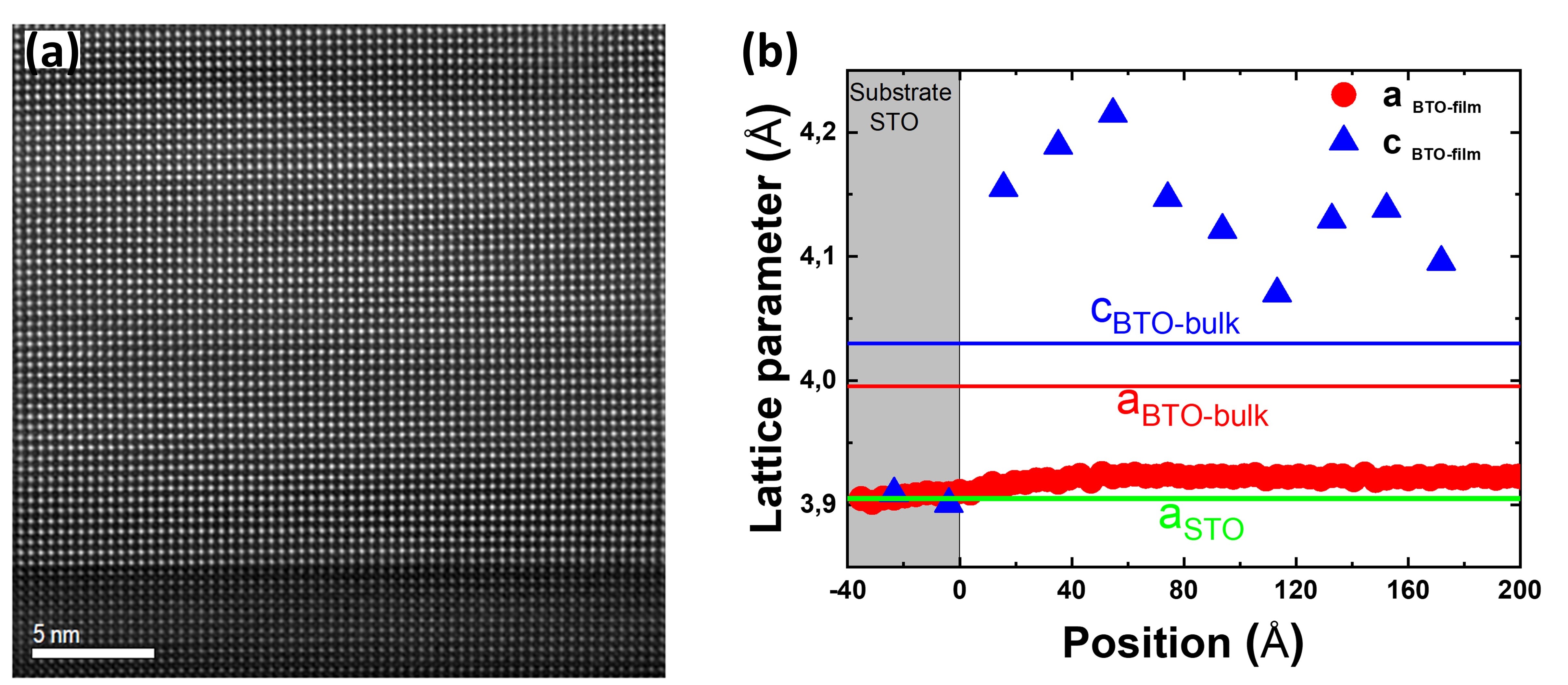}
\caption {(a) Cross sectional HRSTEM-HAADF image of the BTO/STO interface. (b) BTO film lattice parameter profile extracted from HRSTEM-HAADF image.} \label{HR_TEM}
\end{figure}

HRSTEM-HAADF was performed in an image-corrected FEI Titan 80 - 300 keV. A HRSTEM-HAADF image of the BTO/STO interface is presented in Figure \ref{HR_TEM} (a), which shows a well-defined interface between the substrate and the film and an epitaxial growth of the BTO layer. 

The plot in Figure \ref{HR_TEM}(b) illustrates the lattice parameter profile obtained from this image within the first 200 $\AA$.  We estimated a mean $c_{BTO}$ lattice parameter of $4.13\pm \ 0.04 \AA$. The lattice parameter profile shows a smooth relaxation. The structure relaxation of the BTO films depends on the lattice mismatch for BTO films deposited on STO substrates, the lattice mismatch is 2.3$\%$, which is a relatively small value and leads to a smooth relaxation of the structure across the film \cite{kim2021, choi2004enhancement}.  

\begin{figure}[h!]
\includegraphics[width=\textwidth]{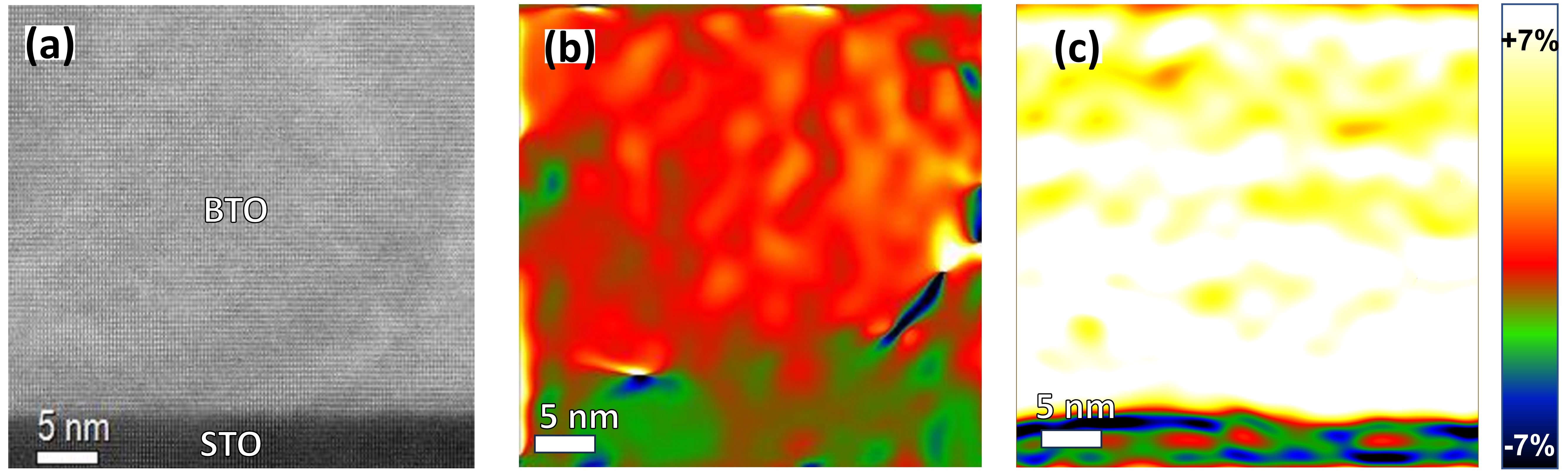}
\caption {(a) BTO film HRSTEM-HAADF image and strain profile obtained from GPA (b) $\epsilon_{xx}$ (c) $\epsilon_{yy}$.} \label{GPA}
\end{figure}

Geometric phase analysis (GPA) was used to obtain the strain profile along the BTO/STO interface. In Figure \ref{GPA} a small compressive strain in the plane of the film can be observed, while the out-of-plane lattice parameter is up to 7\% longer than the bulk value, indicating a tensile strain.

\begin{figure}[ht]
\includegraphics[width=0.7\textwidth]{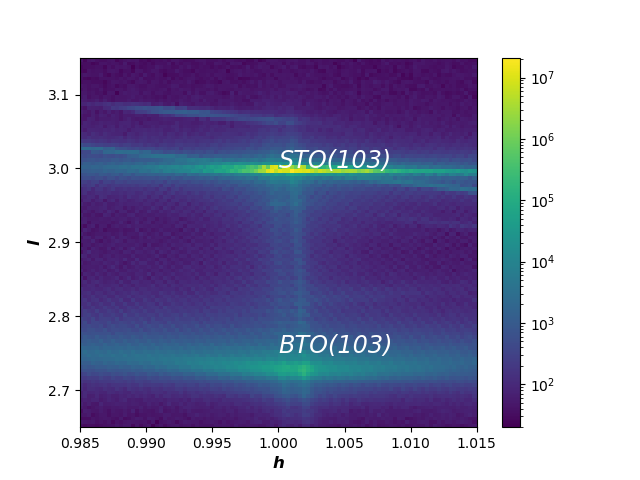}
\caption {RSM around (103) for the 60nm thick BTO layer measured at 150 K.} \label{RSM150K}
\end{figure}

High-resolution synchrotron X-ray diffraction experiments were performed on the XRD2 beamline ($E=$7.00375 keV) at the LNLS (Campinas, Brazil). The samples were cooled down using an Oxford Cryojet5. The temperature dependence of the crystalline structure of BaTiO$_3$ epitaxial films was examined using 2-dimensional HKL-mesh scans in reciprocal space for STO (103) and (113) reflections. These reflections were used as a reference, assuming that the substrate is unstrained, to deduce the BaTiO$_3$ lattice parameters from the corresponding reflections. 

The reciprocal space map of the (103) BTO reflection in the STO $hkl$-mesh was used to evaluate the strain of the structure below room temperature (Figure \ref{RSM150K}). BTO reflection is aligned with the STO peak $h$-coordinate, which indicates that the BTO lattice is clamped to the substrate and therefore strained. The symmetry of the BTO lattice was also analyzed by comparing the $d_{hkl}$, $d_{khl}$, and $d_{-h-kl}$ interplanar distances. The three distances are equal, confirming the tetragonal symmetry of the BTO structures from room temperature to 150 K. The lattice parameters calculated from XRD measurements at room temperature are $a=3.905$ \AA\ and $c=4.286$ \AA, which is strained by 6\% in comparison with the bulk ($c_{BTO}=4.036$ \cite{shebanov1981x}). It should be emphasized that this value represents a mean deformation of the layer, and as shown above, there is a relaxation along the film. 

\paragraph{Ferroelectricity:} The sample was measured using Piezoresponse Force Microscopy to evaluate its ferroelectricity. The measurements were performed at Centro Atómico Bariloche in a Dimension 3100 Brucker microscope using a conductive tip. 

\begin{figure}[ht]
\includegraphics[width=0.7\textwidth]{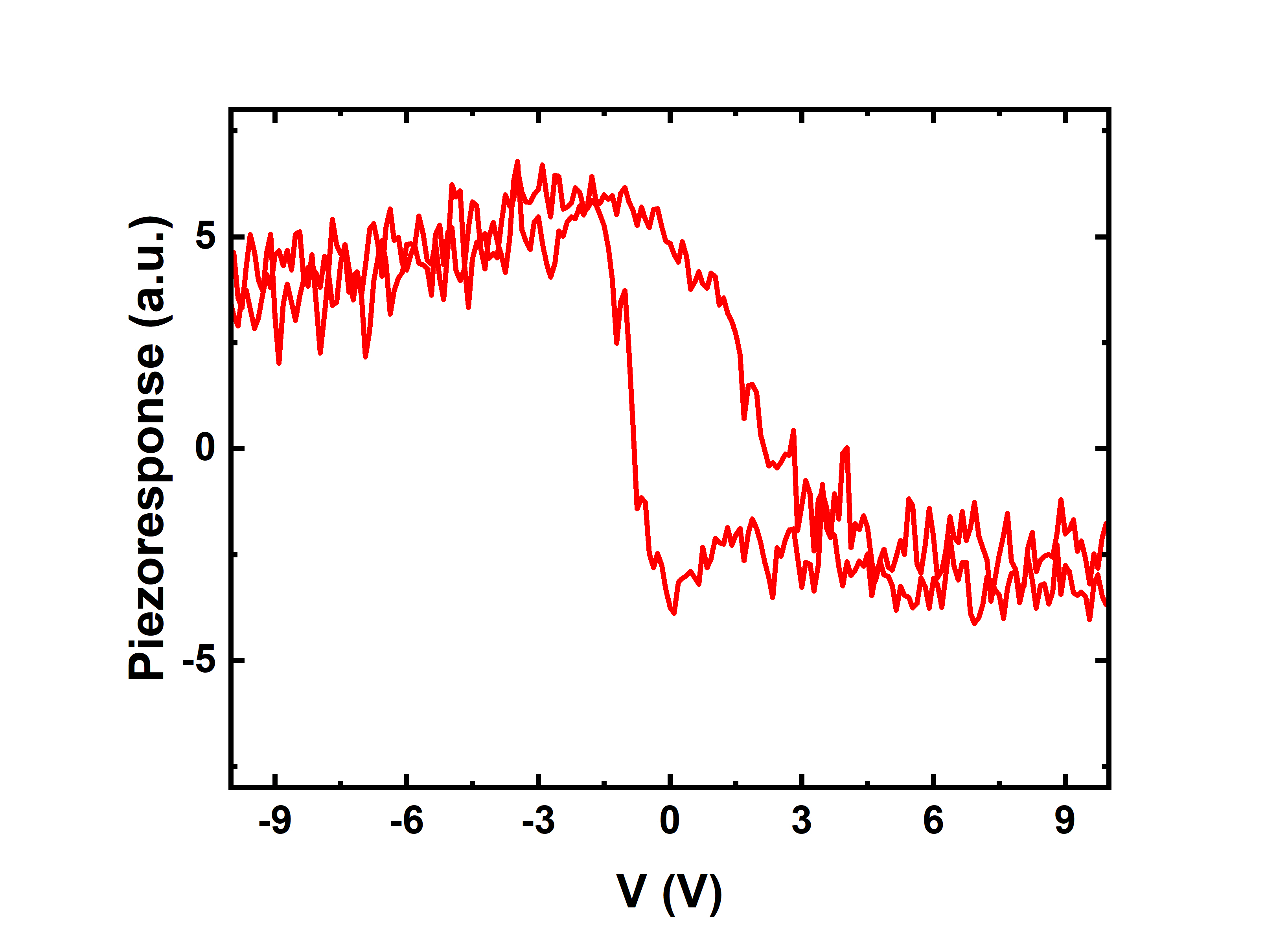}
\caption {Piezoresponse vs voltage bias applied to the conductive tip.} \label{RSM150K}
\end{figure}

The piezoresponse is plotted in Figure~\ref{RSM150K} as a function of the voltage bias applied to the tip. The observed hysteresis in the piezoelectric response confirms the ferroelectric nature of the BTO film \cite{kholkin2007review}.

\bibliography{References}